\documentclass[
 reprint,
superscriptaddress,
 amsmath,amssymb,
prb,
]{revtex4-2}

\usepackage{graphicx}
\usepackage{dcolumn}
\usepackage{bm}

\usepackage[bookmarks=true,colorlinks=true,urlcolor=blue,linkcolor=blue,citecolor=blue,breaklinks=true]{hyperref}
\PassOptionsToPackage{hyphens}{url}
\usepackage{graphicx}
\usepackage[utf8]{inputenc}
\usepackage{amsmath, amssymb}
\usepackage[table]{xcolor}       

\usepackage{lipsum}

\newcommand{\DIPC}[0]{
Donostia International Physics Center (DIPC),
Paseo Manuel de Lardizabal 4, 20018 Donostia-San Sebasti\'an, Spain}
\newcommand{\CFM}[0]{
Centro de F\'{\i}sica de Materiales CFM/MPC (CSIC-UPV/EHU), Paseo Manuel de Lardizabal 5, 20018 Donostia-San Sebasti\'an, Spain}
\newcommand{\PolymerEHU}[0]{Departamento de Pol\'{i}meros y Materiales Avanzados: F\'{i}sica, Qu\'{i}mica y Tecnolog\'{i}a, Facultad de Qu\'{i}micas (UPV/EHU), Apartado 1072, 20080 Donostia-San Sebasti\'{a}n, Spain}

\newcommand{\UPC}[0]{Departament de F\'{i}sica, Universitat Polit\`{e}cnica de Catalunya, Campus Nord B4-B5, E-08034, Barcelona, Spain}

\begin{document}

\title{Anomalous transient blueshift in the internal stretch mode of CO/Pd(111)}

%
\author{Raúl Bombín}
\email{raul.bombin@ehu.eus}
\affiliation{\PolymerEHU}
\affiliation{\CFM}
\affiliation{\UPC}

\author{A. S. Muzas}
\affiliation{\CFM}
\author{Dino Novko}
\affiliation{Institute of Physics, Bijeni\v{c}ka 46, 10000 Zagreb, Croatia}
%
\author{J. I\~{n}aki Juaristi}
\email{josebainaki.juaristi@ehu.eus}
\affiliation{\PolymerEHU}
\affiliation{\CFM}
\affiliation{\DIPC}
\author{Maite Alducin}
\email{maite.alducin@ehu.eus}
\affiliation{\CFM}
\affiliation{\DIPC}

\date{\today}

\begin{abstract}
In time-resolved pump-probe vibrational spectroscopy the internal stretch mode of polar molecules is utilized as a key observable to characterize the ultrafast dynamics of adsorbates on surfaces. The adsorbates non-adiabatic intermode couplings are the commonly accepted mechanisms behind the observed transient frequency shifts. Here, we study the CO/Pd(111) system with a robust theoretical framework that includes electron-hole pair excitations and electron-mediated coupling between the vibrational modes. A mechanism is revealed that screens the electron-phonon interaction and originates a blueshift under ultrafast non-equilibrium conditions. The results are explained in terms of the abrupt change in the density of states around the Fermi level, and are instrumental for understanding dynamics at multi-component surfaces involving localized and standard \emph{s} or \emph{p} states.
\end{abstract}

\maketitle


The use of intense femtosecond laser pulses opened a new and efficient pathway to initiate reactions at surfaces. The interplay between laser induced hot electrons and the nuclear degrees of freedom offers the possibility of activating the adsorbates dynamics within the subpicosecond timescale. As such, femtosecond laser pulses have extensively been used to trigger diverse elementary processes like adsorbates desorption~\cite{budde91,Prybyla1992}, diffusion~\cite{Lawrenz2009}, and different chemical reactions~\cite{frischkorncr06,Park2015}. The main challenge is to understand how the highly excited electrons couple to the adsorbates dynamics and how this energy flows into the different vibrational modes. 

Femtosecond pump-probe vibrational spectroscopy permits tracking in time domain the adsorbates dynamics initiated by the pump laser pulse with unprecedented time-resolution~\cite{arnolds2010}. The internal stretch (IS) vibrational mode of dipolar adsorbates, having a very distinct frequency, appears as the good observable to be followed over the reaction coordinate~\cite{Yampolsky2014,Omiya2014}. Several experiments have been made on Ru, Ir, Pt and Cu surfaces including different coverages of CO and NO molecules~\cite{Bonn2000,Lane2006,Lane2007,Fournier2004,Watanabe2010,Inoue2012,Inoue2016,Omiya2014}. In all those cases, the IS mode frequency exhibits a fast redshift, followed by a slower recovery in the time scale of picoseconds. The frequency shifts are explained within the widely established models in terms of couplings to the low energy CO modes and possible changes in the adsorption site. In both cases, it is the increasing and decreasing population of the CO antibonding 2$\pi^*$ that is thought to cause the weakening and strengthening of the IS mode, respectively. 
An important assumption in these models is 
that the IS mode only brings information on the adsorbates dynamics. However, there are examples showing that there might be other factors contributing to the transient frequency changes. This is the case of the blueshift observed in the rather more complex system formed by CO molecules coordinated to ruthenium tetraphenylporphyrin on a Cu(110) surface~\cite{Omiya2019}, which neither the intermode couplings nor the adsorption site changes are able to explain.

In this Letter we study the ultrafast transient dynamics of 0.5~ML of CO on the Pd(111) surface, using the theoretical framework of Ref.~\cite{Novko2019} that correctly reproduced the experimental ultrafast time-resolved vibrational spectra of CO on Cu(100)~\cite{Inoue2016}. Our calculations predict an unusual blueshift occurring within the subpicosecond time scale that is not related to the commonly accepted intermode coupling, but to the peculiar properties of the surface electronic structure. In solids, the electron-phonon scattering strength is expected to increase with the electronic temperature causing a softening in the frequency because more electrons participate in the interaction~\cite{Baddorf1991,Bracco1996}. This is what explained the subpicosecond redshift in the CO/Cu(100) system~\cite{Novko2019}. Here we show that the strong reduction of the Pd(111) density of states around the Fermi level can revert this normal behavior at the extreme laser-induced high temperatures. Hence the predicted blueshift of the IS mode brings information on the surface electronic structure rather than the intermode couplings. The present results show that dipolar molecules can serve alternatively as a direct probe of ultrafast electron dynamics of metal surfaces, providing time-dependent chemical potential shifts, structure of density of states, electron temperatures, and coupling strengths.

The transient vibrational spectra of the CO adlayer is calculated, following Ref.~\cite{Novko2019}, in terms of the phonon self-energy expressed up to second-order in the electron-phonon (e-ph) interaction, i.e., $\pi_\lambda(\omega, \textbf{q}) \approx \pi_\lambda^{[1]}(\omega,\textbf{q}) + \pi_\lambda^{[2]}(\omega,\textbf{q}$), where $\lambda$, $\omega$, and $\textbf{q}$ denote the index, energy, and momentum of the phonon mode, respectively.
In the femtosecond pump-probe experiments of interest, the CO molecules are initially prepared with their IS mode vibrating in phase by illuminating them with an IR pulse. Thus, only the $\textbf{q}=0$ excitations are considered. 

The expression for the first-order term that exclusively accounts for the electron-hole pair (de)excitations (i.e., the usual nonadiabatic coupling, NC) reads~\cite{Novko2016},
\begin{equation}
\pi_\lambda^{[1]}[\omega; T_e(t)]=\sum_{\mu \mu^\prime\textbf{k}\sigma}\left|g^{\mu\mu^\prime}_\lambda(\textbf{k},0)\right|^2\frac{f(\epsilon_{\mu\textbf{k}})-f(\epsilon_{\mu^\prime\textbf{k}})}{\omega + \epsilon_{\mu\textbf{k}}-\epsilon_{\mu^\prime\textbf{k}}+ i\eta},
\label{eq:first_order}
\end{equation}
where $\mu$, $\textbf{k}$, and $\epsilon_{\mu\textbf{k}}$ are the electron band index, momentum, and energy, respectively; $g^{\mu\mu^\prime}_\lambda(\textbf{k},\textbf{q})$ are the e-ph matrix elements; and the summation over $\sigma$ accounts for the spin degree of freedom. The function $f(\epsilon_{\mu\textbf{k}})=1/(e^{\beta(\epsilon_{\mu\textbf{k}}-\mu(T_e))}+1)$ is the Fermi-Dirac distribution function, where $\beta=1/(k_BT_e)$, $k_B$ is the Boltzmann constant, $T_e$ is the electronic temperature, and $\mu(T_e)$ is the chemical potential at $T_e$.  Following previous works~\cite{Novko2016,Novko2018}, we fix the broadening parameter to a finite, physically motivated value of 30~meV~\cite{Hayashi2013,Schendel2017}. As it will soon become apparent, in the above expression not only the e-ph matrix elements, but also the Pd(111) electronic structure are crucial to understand the ultrafast transient dynamics of the CO adlayer. 

Equation~\eqref{eq:first_order} only contains interband contributions\,\cite{Novko2016}, and the first non-vanishing intraband processes that can affect the CO vibrational dynamics are those involved in the coupling of the IS mode with other phonons. In the framework of many-body perturbation theory, these electron-mediated phonon-phonon coupling (EMPPC) processes are included in the second-order intraband phonon self-energy term that reads~\cite{Novko2018,Giustino2017},
\begin{widetext}
\begin{align}
    \pi_\lambda^{[2]}[\omega; T_e(t), T_l(t)]=
&-\sum_{\mu \mu^\prime\textbf{k}\sigma,\lambda^\prime\textbf{k}^\prime} \left|g^{\mu\mu}_\lambda(\textbf{k},0)\right|^2
\left[1-\frac{g_\lambda^{\mu^\prime\mu^\prime}(\textbf{k}^\prime,0)}{g_\lambda^{\mu\mu}(\textbf{k},0)}\right]
\left|g^{\mu\mu^\prime}_{\lambda^\prime}(\textbf{k},\textbf{q}^\prime)\right|^2\nonumber\\
&\times\sum_{s,s^\prime=\pm 1}
\frac{f(\epsilon_{\mu\textbf{k}})-f(\epsilon_{\mu^\prime\textbf{k}^\prime}-s^\prime s \omega_{\textbf{q}^\prime\lambda^\prime})}
{\epsilon_{\mu\textbf{k}}-(\epsilon_{\mu^\prime\textbf{k}^\prime}-s^\prime s\omega_{\textbf{q}^\prime\lambda^\prime})}
\frac{s\left[n_b(s\omega_{\textbf{q}^\prime\lambda^\prime})+f(s^\prime\epsilon_{\mu^\prime\textbf{k}^\prime})\right]}{\omega\left[\omega + i\eta +s^\prime(\epsilon_{\mu\textbf{k}}-\epsilon_{\mu^\prime\textbf{k}^\prime})+s\omega_{\textbf{q}^\prime\lambda^\prime}\right]},
\label{eq:secon_order}
\end{align}
\end{widetext}
where $\textbf{q}^\prime=\textbf{k}^\prime-\textbf{k}$ and  $n_b(\omega_{\textbf{q},\lambda}) = 1/(e^{\beta\omega_{\textbf{q}\lambda}}-1)$ is the Bose-Einstein distribution, being $\beta=1/(k_BT_l)$ and $T_l$ the lattice temperature. The contribution of the vertex correction [second term in the square bracket of Eq.~\eqref{eq:secon_order}], which is expected to be small~\cite{Bauer1998}, is neglected to reduce the computational cost. Finally, we use the two temperature model (TTM)~\cite{Anisimov1974} to describe the excitation induced by the femtosecond pump pulse in the system as two coupled (non-equilibrated) heat thermal baths. It provides us with the temporal evolution of $T_e$ and $T_l$ that are used to evaluate Eqs.~\eqref{eq:first_order} and~\eqref{eq:secon_order}. 
The transient frequency shift due to e-ph coupling is obtained from the real part of the phonon self-energy as $\Delta\omega(t) \approx \mathrm{Re}[\pi_\lambda(\omega,t)]-\mathrm{Re}[\pi_\lambda(\omega,0)]$. Obviously, the contributions to $\Delta\omega(t)$ from the first- and second-order terms will be $\Delta\omega^{[1]}(t) \approx \mathrm{Re}[\pi_\lambda^{[1]}(\omega,t)]-\mathrm{Re}[\pi_\lambda^{[1]}(\omega,0)$] and  $\Delta\omega^{[2]}(t) \approx \mathrm{Re}[\pi_\lambda^{[2]}(\omega,t)]-\mathrm{Re}[\pi_\lambda^{[2]}(\omega,0)$], respectively.

\begin{figure}
	\centering
	\includegraphics[width=1.00\linewidth]{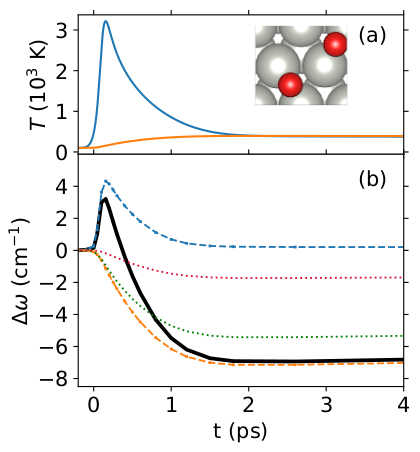}  
\caption{Transient changes induced in CO/Pd(111) by a 450~nm pump pulse (100-fs duration and absorbed fluence of 40~J/m$^2$) that hits the surface at $t=0.1$~ps. The initial temperature is 100~K. (a) Electron $T_e(t)$ (blue) and lattice $T_l(t)$ (orange) temperatures calculated with TTM. Inset: top view of the c(4$\times$2)-2CO$_\textrm{br}$ unit cell. (b) Transient frequency shift of the CO IS mode: $\Delta \omega(t)$ (black solid), $\Delta \omega^{[1]}(t)$ (blue dashed), and $\Delta \omega^{[2]}(t)$ (orange dashed). Contributions of the CO and surface phonons to $\Delta \omega^{[2]}(t)$ are shown by red- and green-dotted lines, respectively.
}
	\label{fig:laser}
\end{figure}

At 0.5~ML, CO adsorbs on Pd(111) with an overall c(4$\times$2) symmetry that consists of two coexisting structures, in which the CO molecules either occupy bridge sites [c(4$\times$2)-2CO$_\textrm{br}$] or the fcc and hcp hollow sites [c(4$\times$2)-2CO$_\textrm{hw}$]~\cite{Rose2002}. The results obtained in both structures for the experimental conditions under consideration are qualitatively similar. Thus, only the results for the c(4$\times$2)-2CO$_\textrm{br}$ structure are discussed here. The inset in Fig.~\ref{fig:laser}(a) shows a top view of the supercell used in these calculations, which includes four Pd-layers with two CO molecules in the adlayer, and 12.9~{\AA} of vacuum.

All the calculations are based on density functional theory and the BEEF-vdW exchange-correlation functional~\cite{Wellendorff2012}, as implemented in the \textsc{Quantum Espresso} package~\cite{Giannozzi2009,Giannozzi2017}. In particular, the electron-phonon Wannier (EPW) code~\cite{NOFFSINGER20102140,PONCE2016116} is used to evaluate the e-ph coupling matrix elements. All the computational details and parameters that are needed to reproduce the results presented here are described in the Supplemental Material~\cite{Suppmat}. 

Figure~\ref{fig:laser}(b) contains the central result of this Letter. The black curve shows that upon exciting the surface with a 450~nm pump laser pulse, the CO IS mode exhibits an unexpected fast initial blueshift that contrast to the usual redshift that has been observed in other metal surfaces by all previous time-resolved pump-probe vibrational spectroscopy experiments~\cite{Bonn2000,Lane2006,Lane2007,Fournier2004,Watanabe2010,Inoue2012,Inoue2016}. The predicted blueshift occurs within the first hundreds of femtoseconds, reaches a maximum value of $\approx$~3~cm$^{-1}$, and progressively vanishes giving rise to a late steady redshift of about $-7$~cm$^{-1}$. The comparison between the time evolution of the transient frequency shift $\Delta\omega (t)$ and that of the electronic and phononic temperatures, $T_e(t)$ and $T_l(t)$ [Fig.~\ref{fig:laser}(a)], suggests that the initial blueshift and subsequent redshift follow $T_e(t)$ and $T_l(t)$, respectively. 

By analyzing the contribution to $\Delta \omega(t)$ coming from the first order interband NC term $\Delta\omega^{[1]}(t)$ and the second-order intraband EMPPC term $\Delta\omega^{[2]}(t)$ [blue- and orange-dashed lines in Fig.~\ref{fig:laser}(b), respectively], it is evident that the blueshift is caused by the coupling to the laser-induced hot electrons. This (positive) NC contribution, which is fully ruled by the time evolution of $T_e$, dominates the IS transient frequency during the initial instants in which $T_e\gg T_l$. The subsequent redshift is caused by the coupling to other system phonons via e-h pairs. This (negative) EMPPC contribution, which clearly increases following $T_l(t)$, dominates the frequency changes as $T_l$ increases above 300~K, being responsible of the almost steady redshift observed at $t\ge$1~ps. Interestingly, a slow steady redshift of comparable magnitude was observed for similar pump laser fluences in CO/Pt(111)~\cite{Watanabe2010}, for which the transient lattice temperatures are expected to be equally large. Figure~\ref{fig:laser}(b) also shows that the EMPPC contribution is dominated by the coupling to the Pd(111) phonons (green-dotted lines) rather than to the rest of CO modes (red-dotted lines). Note that both the surface and the CO adlayer phonons contribute to the EMPPC redshift, discarding any similarity between the present blueshift induced by hot electrons and the blueshift observed under thermal conditions on Ni(111)~\cite{Persson1985} and Pd(100)~\cite{Cook1997}, explained in terms of anharmonic coupling between the CO vibrational modes. 

\begin{figure}[tb!]
	\centering
	\includegraphics[width=1.00\linewidth]{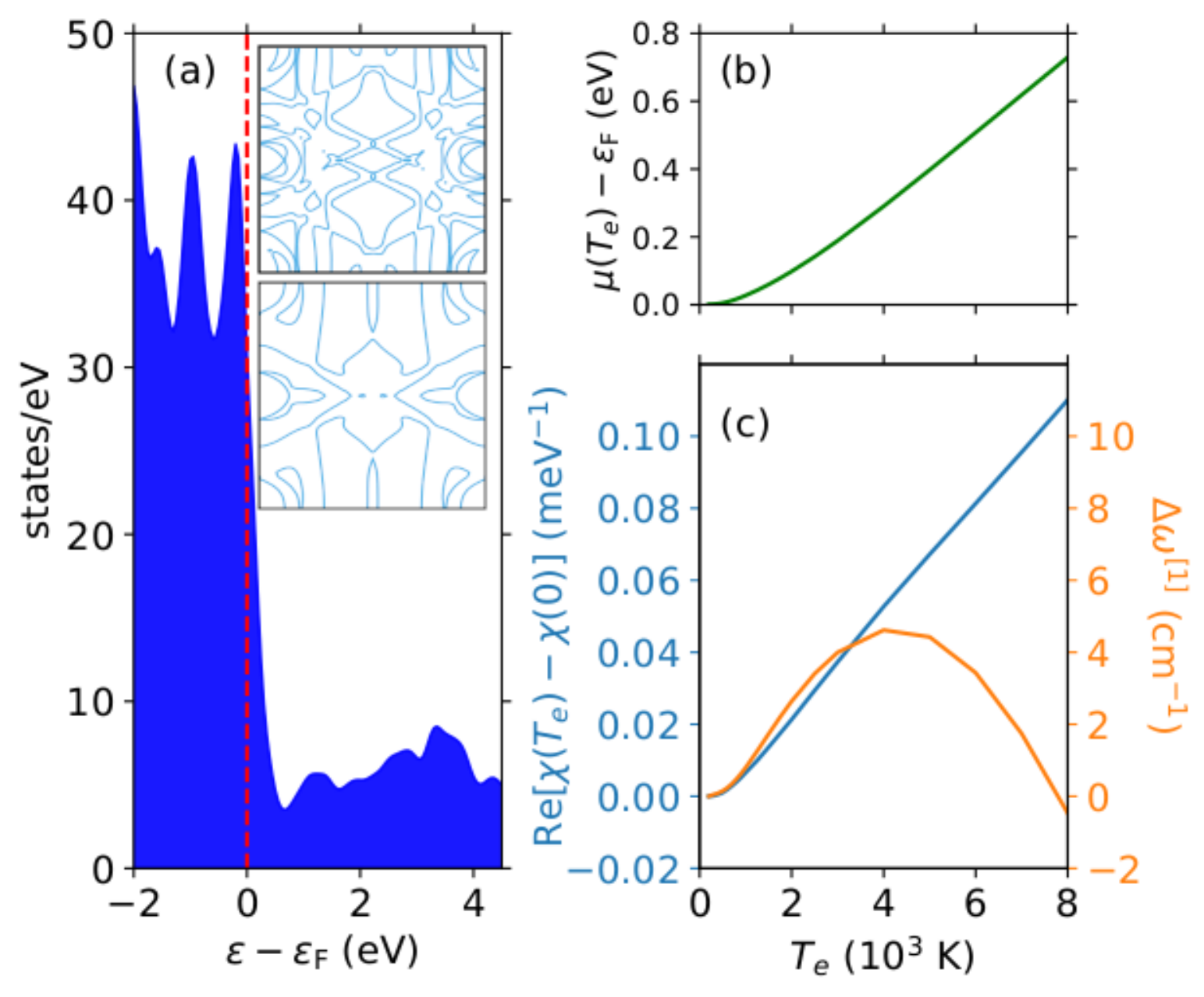}  
\caption{ (a) Electronic DOS of the CO/Pd(111) system. Vertical red dashed line marks the Fermi level. Insets: constant energy cuts of the band structure over the Brillouin zone at $\epsilon=\mu(T_e)$ for $T_e=$~100~K (top) and $T_e=$~3000~K (bottom). (b) Dependence of the chemical potential with the electronic temperature. (c) Change in the real part of $\chi$ (blue) and in $\Delta\omega^{[1]}$ (orange) as a function of $T_e$.
}
	\label{fig:muT}
\end{figure}

The above analysis reveals a different mechanism behind the subpicosend transient blueshift that is correlated to the $T_e$-dependence of the nonadiabatic coupling between the IS mode and the laser-induced hot electrons. One naively should expect an enhancement in the e-ph scattering strength as temperature rises. This is what would cause the transient subpicosecond redshift in the CO/Cu(100) system found by both experiments~\cite{Inoue2016} and theory~\cite{Novko2019}. Thus, the obtained NC blueshift suggesting an effective screening of the e-ph interaction by the hot electrons is somehow anomalous. It must be related to the peculiarities of the CO/Pd(111) electronic structure. Compared to the mentioned CO/Cu(100) system, the presence of the $d$-band edge near $\varepsilon_F$ causes a sharp decrease of the CO/Pd(111) density of states (DOS) around the Fermi level [see Fig.~\ref{fig:muT}(a)]. As $T_e$ increases, the population and depopulation of states above and below $\mu(T_e)$ compensate each other, assuring conservation of the number of electrons $N_e$. Thus, the CO/Pd(111) chemical potential shifts to higher energies to counterbalance the strong asymmetry of the DOS around $\varepsilon_F$. Figure~\ref{fig:muT}(b) shows that the chemical potential, which is calculated numerically by solving iteratively the equality,  $N_e = \int \mathrm{DOS}(\epsilon)f(\epsilon,T_e,\mu(T_e)) d\epsilon$, increases by $0.2$~eV when $T_e$ raises from 100 to 3000~K. Such an increase is enough to induce an important reduction in the DOS around $\mu(T_e)$, i.e., the states that are likely to contribute to the e-ph interaction. To illustrate it, the constant energy cuts of the CO/Pd(111) band structure with the plane $\epsilon=\mu(T_e)$ are shown across the whole Brillouin zone in the insets of Fig.~\ref{fig:muT}(a) for $T_e=$~100~K (top) and $T_e=$~3000~K (bottom). There is an evident reduction of the electronic states that would support the above interpretation of the blueshift as a weakening of the e-ph interaction. The results calculated with $\mu(T_e) = \varepsilon_F$ further confirm this idea~\cite{Suppmat}.

The dependence of the NC term $\Delta \omega^{[1]}$ on $T_e$ is not monotonic. As shown in Fig.~\ref{fig:muT}, $\Delta \omega^{[1]} (T_e)$ (orange curve, right axis) initially increases with $T_e$ up to reaching its maximum value at around 4000~K and it starts to decrease, becoming negative at temperatures larger than 7000~K. There are two factors ruling the first-order phonon self-energy $\pi^{[1]}$ that can affect the temperature dependence of $\Delta \omega^{[1]}$, the e-ph matrix elements and the electronic structure. To disentangle each contribution, it is useful to evaluate the response function $\chi(\omega)$ under the constant matrix elements approximation~\cite{SkyZhou2020}, that reads,
\begin{equation}
\chi(\omega)=\sum_{\mu \mu^\prime\textbf{k}\sigma}
\frac{f(\epsilon_{\mu\textbf{k}})-f(\epsilon_{\mu^\prime\textbf{k}})}{\omega + \epsilon_{\mu\textbf{k}}-\epsilon_{\mu^\prime\textbf{k}}+ i\eta}.
\label{eq:response}
\end{equation}
Mathematically, the above expression coincides with setting the e-ph matrix elements to one in the first-order phonon self-energy expression [\textit{cf}. Eq.~\eqref{eq:first_order}], allowing us to single out the effect of the electronic structure in the temperature dependence of $\pi^{[1]}(\omega,T_e)$. Figure~\ref{fig:muT}(c) shows that the change in the real part of the response function (blue, left axis) grows monotonically with $T_e$ in contrast to the non-monotonous behavior of $\Delta \omega^{[1]}$. Altogether, these results suggest that the initial blueshift that is observed in the initial stage dynamics of Fig.~\ref{fig:laser}(b) is a direct consequence of the electronic structure. This mechanism competes with the strength of the e-ph coupling that tends to redshift the frequency as $T_e$ increases. It is only for extreme large electronic temperatures that the latter will dominate the frequency change. Therefore, the next step is to elucidate how the amount of energy that is deposited on the system affects the IS mode dynamics.

\begin{figure}
	\centering
	\includegraphics[width=1.00\linewidth]{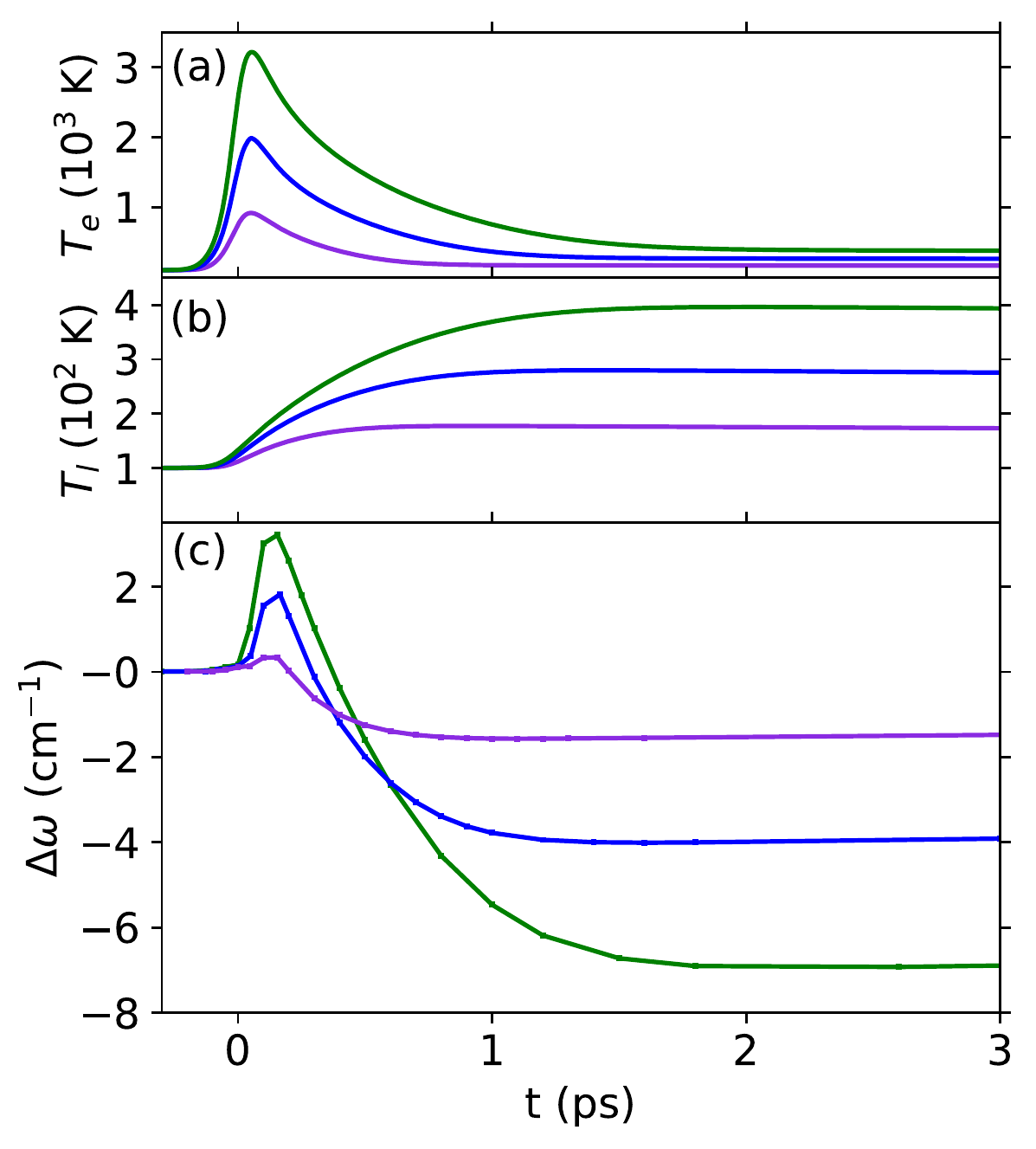}  
\caption{Transient changes induced in CO/Pd(111) by a 450~nm pump pulse (100-fs duration) with different absorbed fluences: $F=$~6 (purple lines), 19 (blue lines), and 40~J/m$^2$ (green lines). The pulse maximum is at $t=0.1$~ps and the system initial temperature is 100~K.  (a) Electron temperature $T_e(t)$, (b) lattice temperature $T_l(t)$, and (c) transient frequency shift of the CO IS mode $\Delta \omega(t)$.
}
	\label{fig:fluences}
\end{figure}

We compare in Fig.~\ref{fig:fluences} the results obtained for three different pump laser absorbed fluences: $F =$~6, 19, and 40~J/m$^2$. Figure~\ref{fig:fluences}~(c) shows that the behavior of the transient frequency shift $\Delta \omega (t)$ is qualitatively similar for the three absorbed fluences. Both the fast initial blueshift and the subsequent steady redshift increase in magnitude as $F$ increases. In particular, for the lowest fluence of 6~J/m$^2$, the highest $T_e$ values of about 1000~K [see Fig.~\ref{fig:fluences}(a)] are clearly insufficient to induce a sizable blueshift in the frequency. However, we expect it to be experimentally accessible for fluences as low as 19~J/m$^2$, provided that $T_e$ temporally reaches values of $\approx$~2000K, as shown in Fig.~\ref{fig:fluences}(a). Let us finally remark that in experiments the pump pulse creates a non-equilibrium electron distribution during the first tenths of femtoseconds that is not included in our calculations. This could introduce a competing mechanism that eventually may affect the initial transient blueshift. However, considering that the inelastic lifetime of electrons in Pd (around $10$~fs~\cite{Zhukov2002}) is much smaller than the width of the pump pulse (100~fs) and that thermalization occurs continuously, the non-equilibrated electrons in this system are expected to thermalize very rapidly without masking the predicted blueshift.

In summary, our calculations reveal that the e-ph interaction can be screened upon femtosecond laser pulse irradiation provided the electronic density of states of the system changes abruptly around the Fermi level. In particular, we study the ultrafast transient dynamics of the internal stretch mode of CO adsorbed on Pd(111). To do so, we combine density functional theory with many-body perturbation theory and evaluate the phonon self-energy up to second order in the e-ph interaction~\cite{Novko2019}. This allows us to characterize the two main mechanisms that participate in the vibrational relaxation of the internal stretch mode: the first order interband non-adiabatic coupling and the intraband electron mediated phonon-phonon coupling. Under current state-of-the-art femtosecond pump-probe experimental conditions, the large transient electronic temperatures that are induced in the femtosecond regime (with $T_e\gg T_l$) give rise to an unconventional blueshift in its frequency.  The latter is followed in the picosecond regime by a redshift, larger in magnitude. We show that the initial fast blueshift arises purely from temperature-dependent electronic structure effects, while the subsequent redshift occurs due to the coupling of the internal stretch mode with other phononic modes. 
The similarity of the results presented here and those of Ref.~\cite{Omiya2019} suggests that an analogous mechanism might be behind the blueshift that was observed for CO molecules coordinated to Ruthenium tetraphenylporphyrin on a Cu(110), where large planar molecules might introduce localized high-density states close to Fermi level in equivalence to the \emph{d} states in the present case.  Our results are accessible to current experimental setups and we hope they will stimulate further research on the ultrafast dynamics of polar molecules on complex metallic surfaces with localized states.

The authors acknowledge financial support by the Gobierno Vasco-UPV/EHU
Project No. IT1246-19 and the Spanish Ministerio de Ciencia e Innovación [Grant No. PID2019-107396GB-I00/AEI/10.13039/501100011033]. R. B. also acknowledges European Union-NextGenerationEU, Ministry of Universities and Recovery, Transformation and Resilience Plan, through a call from Polytechnic University of Catalonia. D.N. additionally acknowledges financial support from the Croatian Science Foundation (Grant no. UIP-2019-04-6869). This research was conducted in the scope of the Transnational Common Laboratory (LTC) “QuantumChemPhys – Theoretical Chemistry and Physics at the Quantum Scale”.  Computational resources were provided by the DIPC computing center. 

\bibliography{refs}

\begin{thebibliography}{28}%
\makeatletter
\providecommand \@ifxundefined [1]{%
 \@ifx{#1\undefined}
}%
\providecommand \@ifnum [1]{%
 \ifnum #1\expandafter \@firstoftwo
 \else \expandafter \@secondoftwo
 \fi
}%
\providecommand \@ifx [1]{%
 \ifx #1\expandafter \@firstoftwo
 \else \expandafter \@secondoftwo
 \fi
}%
\providecommand \natexlab [1]{#1}%
\providecommand \enquote  [1]{``#1''}%
\providecommand \bibnamefont  [1]{#1}%
\providecommand \bibfnamefont [1]{#1}%
\providecommand \citenamefont [1]{#1}%
\providecommand \href@noop [0]{\@secondoftwo}%
\providecommand \href [0]{\begingroup \@sanitize@url \@href}%
\providecommand \@href[1]{\@@startlink{#1}\@@href}%
\providecommand \@@href[1]{\endgroup#1\@@endlink}%
\providecommand \@sanitize@url [0]{\catcode `\\12\catcode `\$12\catcode
  `\&12\catcode `\#12\catcode `\^12\catcode `\_12\catcode `\%12\relax}%
\providecommand \@@startlink[1]{}%
\providecommand \@@endlink[0]{}%
\providecommand \url  [0]{\begingroup\@sanitize@url \@url }%
\providecommand \@url [1]{\endgroup\@href {#1}{\urlprefix }}%
\providecommand \urlprefix  [0]{URL }%
\providecommand \Eprint [0]{\href }%
\providecommand \doibase [0]{https://doi.org/}%
\providecommand \selectlanguage [0]{\@gobble}%
\providecommand \bibinfo  [0]{\@secondoftwo}%
\providecommand \bibfield  [0]{\@secondoftwo}%
\providecommand \translation [1]{[#1]}%
\providecommand \BibitemOpen [0]{}%
\providecommand \bibitemStop [0]{}%
\providecommand \bibitemNoStop [0]{.\EOS\space}%
\providecommand \EOS [0]{\spacefactor3000\relax}%
\providecommand \BibitemShut  [1]{\csname bibitem#1\endcsname}%
\let\auto@bib@innerbib\@empty
\bibitem [{\citenamefont {Giannozzi}\ \emph {et~al.}(2009)\citenamefont
  {Giannozzi}, \citenamefont {Baroni}, \citenamefont {Bonini}, \citenamefont
  {Calandra}, \citenamefont {Car}, \citenamefont {Cavazzoni}, \citenamefont
  {Ceresoli}, \citenamefont {Chiarotti}, \citenamefont {Cococcioni},
  \citenamefont {Dabo}, \citenamefont {Dal~Corso}, \citenamefont
  {De~Gironcoli}, \citenamefont {Fabris}, \citenamefont {Fratesi},
  \citenamefont {Gebauer}, \citenamefont {Gerstmann}, \citenamefont
  {Gougoussis}, \citenamefont {Kokalj}, \citenamefont {Lazzeri}, \citenamefont
  {Martin-Samos}, \citenamefont {Marzari}, \citenamefont {Mauri}, \citenamefont
  {Mazzarello}, \citenamefont {Paolini}, \citenamefont {Pasquarello},
  \citenamefont {Paulatto}, \citenamefont {Sbraccia}, \citenamefont {Scandolo},
  \citenamefont {Sclauzero}, \citenamefont {Seitsonen}, \citenamefont
  {Smogunov}, \citenamefont {Umari},\ and\ \citenamefont
  {Wentzcovitch}}]{Giannozzi2009}%
  \BibitemOpen
  \bibfield  {author} {\bibinfo {author} {\bibfnamefont {P.}~\bibnamefont
  {Giannozzi}}, \bibinfo {author} {\bibfnamefont {S.}~\bibnamefont {Baroni}},
  \bibinfo {author} {\bibfnamefont {N.}~\bibnamefont {Bonini}}, \bibinfo
  {author} {\bibfnamefont {M.}~\bibnamefont {Calandra}}, \bibinfo {author}
  {\bibfnamefont {R.}~\bibnamefont {Car}}, \bibinfo {author} {\bibfnamefont
  {C.}~\bibnamefont {Cavazzoni}}, \bibinfo {author} {\bibfnamefont
  {D.}~\bibnamefont {Ceresoli}}, \bibinfo {author} {\bibfnamefont {G.~L.}\
  \bibnamefont {Chiarotti}}, \bibinfo {author} {\bibfnamefont {M.}~\bibnamefont
  {Cococcioni}}, \bibinfo {author} {\bibfnamefont {I.}~\bibnamefont {Dabo}},
  \bibinfo {author} {\bibfnamefont {A.}~\bibnamefont {Dal~Corso}}, \bibinfo
  {author} {\bibfnamefont {S.}~\bibnamefont {De~Gironcoli}}, \bibinfo {author}
  {\bibfnamefont {S.}~\bibnamefont {Fabris}}, \bibinfo {author} {\bibfnamefont
  {G.}~\bibnamefont {Fratesi}}, \bibinfo {author} {\bibfnamefont
  {R.}~\bibnamefont {Gebauer}}, \bibinfo {author} {\bibfnamefont
  {U.}~\bibnamefont {Gerstmann}}, \bibinfo {author} {\bibfnamefont
  {C.}~\bibnamefont {Gougoussis}}, \bibinfo {author} {\bibfnamefont
  {A.}~\bibnamefont {Kokalj}}, \bibinfo {author} {\bibfnamefont
  {M.}~\bibnamefont {Lazzeri}}, \bibinfo {author} {\bibfnamefont
  {L.}~\bibnamefont {Martin-Samos}}, \bibinfo {author} {\bibfnamefont
  {N.}~\bibnamefont {Marzari}}, \bibinfo {author} {\bibfnamefont
  {F.}~\bibnamefont {Mauri}}, \bibinfo {author} {\bibfnamefont
  {R.}~\bibnamefont {Mazzarello}}, \bibinfo {author} {\bibfnamefont
  {S.}~\bibnamefont {Paolini}}, \bibinfo {author} {\bibfnamefont
  {A.}~\bibnamefont {Pasquarello}}, \bibinfo {author} {\bibfnamefont
  {L.}~\bibnamefont {Paulatto}}, \bibinfo {author} {\bibfnamefont
  {C.}~\bibnamefont {Sbraccia}}, \bibinfo {author} {\bibfnamefont
  {S.}~\bibnamefont {Scandolo}}, \bibinfo {author} {\bibfnamefont
  {G.}~\bibnamefont {Sclauzero}}, \bibinfo {author} {\bibfnamefont {A.~P.}\
  \bibnamefont {Seitsonen}}, \bibinfo {author} {\bibfnamefont {A.}~\bibnamefont
  {Smogunov}}, \bibinfo {author} {\bibfnamefont {P.}~\bibnamefont {Umari}},\
  and\ \bibinfo {author} {\bibfnamefont {R.~M.}\ \bibnamefont {Wentzcovitch}},\
  }\bibfield  {title} {\bibinfo {title} {{QUANTUM ESPRESSO: A modular and
  open-source software project for quantum simulations of materials}},\ }\href
  {https://doi.org/10.1088/0953-8984/21/39/395502} {\bibfield  {journal}
  {\bibinfo  {journal} {J. Phys. Condens. Matter}\ }\textbf {\bibinfo {volume}
  {21}},\ \bibinfo {pages} {395502} (\bibinfo {year} {2009})}\BibitemShut
  {NoStop}%
\bibitem [{\citenamefont {Giannozzi}\ \emph {et~al.}(2017)\citenamefont
  {Giannozzi}, \citenamefont {Andreussi}, \citenamefont {Brumme}, \citenamefont
  {Bunau}, \citenamefont {{Buongiorno Nardelli}}, \citenamefont {Calandra},
  \citenamefont {Car}, \citenamefont {Cavazzoni}, \citenamefont {Ceresoli},
  \citenamefont {Cococcioni}, \citenamefont {Colonna}, \citenamefont
  {Carnimeo}, \citenamefont {{Dal Corso}}, \citenamefont {{De Gironcoli}},
  \citenamefont {Delugas}, \citenamefont {Distasio}, \citenamefont {Ferretti},
  \citenamefont {Floris}, \citenamefont {Fratesi}, \citenamefont {Fugallo},
  \citenamefont {Gebauer}, \citenamefont {Gerstmann}, \citenamefont {Giustino},
  \citenamefont {Gorni}, \citenamefont {Jia}, \citenamefont {Kawamura},
  \citenamefont {Ko}, \citenamefont {Kokalj}, \citenamefont
  {K{\"{u}}c{\"{u}}kbenli}, \citenamefont {Lazzeri}, \citenamefont {Marsili},
  \citenamefont {Marzari}, \citenamefont {Mauri}, \citenamefont {Nguyen},
  \citenamefont {Nguyen}, \citenamefont {Otero-De-La-Roza}, \citenamefont
  {Paulatto}, \citenamefont {Ponc{\'{e}}}, \citenamefont {Rocca}, \citenamefont
  {Sabatini}, \citenamefont {Santra}, \citenamefont {Schlipf}, \citenamefont
  {Seitsonen}, \citenamefont {Smogunov}, \citenamefont {Timrov}, \citenamefont
  {Thonhauser}, \citenamefont {Umari}, \citenamefont {Vast}, \citenamefont
  {Wu},\ and\ \citenamefont {Baroni}}]{Giannozzi2017}%
  \BibitemOpen
  \bibfield  {author} {\bibinfo {author} {\bibfnamefont {P.}~\bibnamefont
  {Giannozzi}}, \bibinfo {author} {\bibfnamefont {O.}~\bibnamefont
  {Andreussi}}, \bibinfo {author} {\bibfnamefont {T.}~\bibnamefont {Brumme}},
  \bibinfo {author} {\bibfnamefont {O.}~\bibnamefont {Bunau}}, \bibinfo
  {author} {\bibfnamefont {M.}~\bibnamefont {{Buongiorno Nardelli}}}, \bibinfo
  {author} {\bibfnamefont {M.}~\bibnamefont {Calandra}}, \bibinfo {author}
  {\bibfnamefont {R.}~\bibnamefont {Car}}, \bibinfo {author} {\bibfnamefont
  {C.}~\bibnamefont {Cavazzoni}}, \bibinfo {author} {\bibfnamefont
  {D.}~\bibnamefont {Ceresoli}}, \bibinfo {author} {\bibfnamefont
  {M.}~\bibnamefont {Cococcioni}}, \bibinfo {author} {\bibfnamefont
  {N.}~\bibnamefont {Colonna}}, \bibinfo {author} {\bibfnamefont
  {I.}~\bibnamefont {Carnimeo}}, \bibinfo {author} {\bibfnamefont
  {A.}~\bibnamefont {{Dal Corso}}}, \bibinfo {author} {\bibfnamefont
  {S.}~\bibnamefont {{De Gironcoli}}}, \bibinfo {author} {\bibfnamefont
  {P.}~\bibnamefont {Delugas}}, \bibinfo {author} {\bibfnamefont {R.~A.}\
  \bibnamefont {Distasio}}, \bibinfo {author} {\bibfnamefont {A.}~\bibnamefont
  {Ferretti}}, \bibinfo {author} {\bibfnamefont {A.}~\bibnamefont {Floris}},
  \bibinfo {author} {\bibfnamefont {G.}~\bibnamefont {Fratesi}}, \bibinfo
  {author} {\bibfnamefont {G.}~\bibnamefont {Fugallo}}, \bibinfo {author}
  {\bibfnamefont {R.}~\bibnamefont {Gebauer}}, \bibinfo {author} {\bibfnamefont
  {U.}~\bibnamefont {Gerstmann}}, \bibinfo {author} {\bibfnamefont
  {F.}~\bibnamefont {Giustino}}, \bibinfo {author} {\bibfnamefont
  {T.}~\bibnamefont {Gorni}}, \bibinfo {author} {\bibfnamefont
  {J.}~\bibnamefont {Jia}}, \bibinfo {author} {\bibfnamefont {M.}~\bibnamefont
  {Kawamura}}, \bibinfo {author} {\bibfnamefont {H.~Y.}\ \bibnamefont {Ko}},
  \bibinfo {author} {\bibfnamefont {A.}~\bibnamefont {Kokalj}}, \bibinfo
  {author} {\bibfnamefont {E.}~\bibnamefont {K{\"{u}}c{\"{u}}kbenli}}, \bibinfo
  {author} {\bibfnamefont {M.}~\bibnamefont {Lazzeri}}, \bibinfo {author}
  {\bibfnamefont {M.}~\bibnamefont {Marsili}}, \bibinfo {author} {\bibfnamefont
  {N.}~\bibnamefont {Marzari}}, \bibinfo {author} {\bibfnamefont
  {F.}~\bibnamefont {Mauri}}, \bibinfo {author} {\bibfnamefont {N.~L.}\
  \bibnamefont {Nguyen}}, \bibinfo {author} {\bibfnamefont {H.~V.}\
  \bibnamefont {Nguyen}}, \bibinfo {author} {\bibfnamefont {A.}~\bibnamefont
  {Otero-De-La-Roza}}, \bibinfo {author} {\bibfnamefont {L.}~\bibnamefont
  {Paulatto}}, \bibinfo {author} {\bibfnamefont {S.}~\bibnamefont
  {Ponc{\'{e}}}}, \bibinfo {author} {\bibfnamefont {D.}~\bibnamefont {Rocca}},
  \bibinfo {author} {\bibfnamefont {R.}~\bibnamefont {Sabatini}}, \bibinfo
  {author} {\bibfnamefont {B.}~\bibnamefont {Santra}}, \bibinfo {author}
  {\bibfnamefont {M.}~\bibnamefont {Schlipf}}, \bibinfo {author} {\bibfnamefont
  {A.~P.}\ \bibnamefont {Seitsonen}}, \bibinfo {author} {\bibfnamefont
  {A.}~\bibnamefont {Smogunov}}, \bibinfo {author} {\bibfnamefont
  {I.}~\bibnamefont {Timrov}}, \bibinfo {author} {\bibfnamefont
  {T.}~\bibnamefont {Thonhauser}}, \bibinfo {author} {\bibfnamefont
  {P.}~\bibnamefont {Umari}}, \bibinfo {author} {\bibfnamefont
  {N.}~\bibnamefont {Vast}}, \bibinfo {author} {\bibfnamefont {X.}~\bibnamefont
  {Wu}},\ and\ \bibinfo {author} {\bibfnamefont {S.}~\bibnamefont {Baroni}},\
  }\bibfield  {title} {\bibinfo {title} {{Advanced} capabilities for materials
  modelling with {Quantum ESPRESSO}},\ }\href
  {https://doi.org/10.1088/1361-648X/aa8f79} {\bibfield  {journal} {\bibinfo
  {journal} {J. Phys. Condens. Matter}\ }\textbf {\bibinfo {volume} {29}},\
  \bibinfo {pages} {465901} (\bibinfo {year} {2017})}\BibitemShut {NoStop}%
\bibitem [{\citenamefont {Wellendorff}\ \emph {et~al.}(2012)\citenamefont
  {Wellendorff}, \citenamefont {Lundgaard}, \citenamefont {M\o{}gelh\o{}j},
  \citenamefont {Petzold}, \citenamefont {Landis}, \citenamefont {N\o{}rskov},
  \citenamefont {Bligaard},\ and\ \citenamefont {Jacobsen}}]{Wellendorff2012}%
  \BibitemOpen
  \bibfield  {author} {\bibinfo {author} {\bibfnamefont {J.}~\bibnamefont
  {Wellendorff}}, \bibinfo {author} {\bibfnamefont {K.~T.}\ \bibnamefont
  {Lundgaard}}, \bibinfo {author} {\bibfnamefont {A.}~\bibnamefont
  {M\o{}gelh\o{}j}}, \bibinfo {author} {\bibfnamefont {V.}~\bibnamefont
  {Petzold}}, \bibinfo {author} {\bibfnamefont {D.~D.}\ \bibnamefont {Landis}},
  \bibinfo {author} {\bibfnamefont {J.~K.}\ \bibnamefont {N\o{}rskov}},
  \bibinfo {author} {\bibfnamefont {T.}~\bibnamefont {Bligaard}},\ and\
  \bibinfo {author} {\bibfnamefont {K.~W.}\ \bibnamefont {Jacobsen}},\
  }\bibfield  {title} {\bibinfo {title} {Density functionals for surface
  science: Exchange-correlation model development with bayesian error
  estimation},\ }\href {https://doi.org/10.1103/PhysRevB.85.235149} {\bibfield
  {journal} {\bibinfo  {journal} {Phys. Rev. B}\ }\textbf {\bibinfo {volume}
  {85}},\ \bibinfo {pages} {235149} (\bibinfo {year} {2012})}\BibitemShut
  {NoStop}%
\bibitem [{\citenamefont {{Dal Corso}}(2014{\natexlab{a}})}]{dalcorso2014}%
  \BibitemOpen
  \bibfield  {author} {\bibinfo {author} {\bibfnamefont {A.}~\bibnamefont {{Dal
  Corso}}},\ }\bibfield  {title} {\bibinfo {title} {Pseudopotentials periodic
  table: From {H} to {Pu}},\ }\href
  {https://doi.org/https://doi.org/10.1016/j.commatsci.2014.07.043} {\bibfield
  {journal} {\bibinfo  {journal} {Comp. Mater. Sci.}\ }\textbf {\bibinfo
  {volume} {95}},\ \bibinfo {pages} {337} (\bibinfo {year}
  {2014}{\natexlab{a}})}\BibitemShut {NoStop}%
\bibitem [{\citenamefont {{Dal Corso}}(2014{\natexlab{b}})}]{dalcorsolink}%
  \BibitemOpen
  \bibfield  {author} {\bibinfo {author} {\bibfnamefont {A.}~\bibnamefont {{Dal
  Corso}}},\ }\href@noop {} {}\bibinfo {howpublished}
  {\url{https://dalcorso.github.io/pslibrary/}} (\bibinfo {year}
  {2014}{\natexlab{b}})\BibitemShut {NoStop}%
\bibitem [{\citenamefont {Monkhorst}\ and\ \citenamefont
  {Pack}(1976)}]{Monkhorst1976}%
  \BibitemOpen
  \bibfield  {author} {\bibinfo {author} {\bibfnamefont {H.~J.}\ \bibnamefont
  {Monkhorst}}\ and\ \bibinfo {author} {\bibfnamefont {J.~D.}\ \bibnamefont
  {Pack}},\ }\bibfield  {title} {\bibinfo {title} {Special points for
  brillouin-zone integrations},\ }\href
  {https://doi.org/10.1103/PhysRevB.13.5188} {\bibfield  {journal} {\bibinfo
  {journal} {Phys. Rev. B}\ }\textbf {\bibinfo {volume} {13}},\ \bibinfo
  {pages} {5188} (\bibinfo {year} {1976})}\BibitemShut {NoStop}%
\bibitem [{\citenamefont {Noffsinger}\ \emph {et~al.}(2010)\citenamefont
  {Noffsinger}, \citenamefont {Giustino}, \citenamefont {Malone}, \citenamefont
  {Park}, \citenamefont {Louie},\ and\ \citenamefont
  {Cohen}}]{NOFFSINGER20102140}%
  \BibitemOpen
  \bibfield  {author} {\bibinfo {author} {\bibfnamefont {J.}~\bibnamefont
  {Noffsinger}}, \bibinfo {author} {\bibfnamefont {F.}~\bibnamefont
  {Giustino}}, \bibinfo {author} {\bibfnamefont {B.~D.}\ \bibnamefont
  {Malone}}, \bibinfo {author} {\bibfnamefont {C.-H.}\ \bibnamefont {Park}},
  \bibinfo {author} {\bibfnamefont {S.~G.}\ \bibnamefont {Louie}},\ and\
  \bibinfo {author} {\bibfnamefont {M.~L.}\ \bibnamefont {Cohen}},\ }\bibfield
  {title} {\bibinfo {title} {Epw: A program for calculating the
  electron–phonon coupling using maximally localized {Wannier} functions},\
  }\href {https://doi.org/https://doi.org/10.1016/j.cpc.2010.08.027} {\bibfield
   {journal} {\bibinfo  {journal} {Comput. Phys. Commun.}\ }\textbf {\bibinfo
  {volume} {181}},\ \bibinfo {pages} {2140} (\bibinfo {year}
  {2010})}\BibitemShut {NoStop}%
\bibitem [{\citenamefont {Poncé}\ \emph {et~al.}(2016)\citenamefont {Poncé},
  \citenamefont {Margine}, \citenamefont {Verdi},\ and\ \citenamefont
  {Giustino}}]{PONCE2016116}%
  \BibitemOpen
  \bibfield  {author} {\bibinfo {author} {\bibfnamefont {S.}~\bibnamefont
  {Poncé}}, \bibinfo {author} {\bibfnamefont {E.}~\bibnamefont {Margine}},
  \bibinfo {author} {\bibfnamefont {C.}~\bibnamefont {Verdi}},\ and\ \bibinfo
  {author} {\bibfnamefont {F.}~\bibnamefont {Giustino}},\ }\bibfield  {title}
  {\bibinfo {title} {Epw: Electron–phonon coupling, transport and
  superconducting properties using maximally localized {Wannier} functions},\
  }\href {https://doi.org/https://doi.org/10.1016/j.cpc.2016.07.028} {\bibfield
   {journal} {\bibinfo  {journal} {Comput. Phys. Commun.}\ }\textbf {\bibinfo
  {volume} {209}},\ \bibinfo {pages} {116 } (\bibinfo {year}
  {2016})}\BibitemShut {NoStop}%
\bibitem [{\citenamefont {Vitale}\ \emph {et~al.}(2020)\citenamefont {Vitale},
  \citenamefont {Pizzi}, \citenamefont {Marrazzo}, \citenamefont {Yates},
  \citenamefont {Marzari},\ and\ \citenamefont {Mostofi}}]{Vitale2020}%
  \BibitemOpen
  \bibfield  {author} {\bibinfo {author} {\bibfnamefont {V.}~\bibnamefont
  {Vitale}}, \bibinfo {author} {\bibfnamefont {G.}~\bibnamefont {Pizzi}},
  \bibinfo {author} {\bibfnamefont {A.}~\bibnamefont {Marrazzo}}, \bibinfo
  {author} {\bibfnamefont {J.~R.}\ \bibnamefont {Yates}}, \bibinfo {author}
  {\bibfnamefont {N.}~\bibnamefont {Marzari}},\ and\ \bibinfo {author}
  {\bibfnamefont {A.~A.}\ \bibnamefont {Mostofi}},\ }\bibfield  {title}
  {\bibinfo {title} {Automated high-throughput wannierisation},\ }\href
  {https://doi.org/10.1038/s41524-020-0312-y} {\bibfield  {journal} {\bibinfo
  {journal} {npj Computational Materials}\ }\textbf {\bibinfo {volume} {6}},\
  \bibinfo {pages} {66} (\bibinfo {year} {2020})}\BibitemShut {NoStop}%
\bibitem [{\citenamefont {Mostofi}\ \emph {et~al.}(2008)\citenamefont
  {Mostofi}, \citenamefont {Yates}, \citenamefont {Lee}, \citenamefont {Souza},
  \citenamefont {Vanderbilt},\ and\ \citenamefont {Marzari}}]{Mostofi2008}%
  \BibitemOpen
  \bibfield  {author} {\bibinfo {author} {\bibfnamefont {A.~A.}\ \bibnamefont
  {Mostofi}}, \bibinfo {author} {\bibfnamefont {J.~R.}\ \bibnamefont {Yates}},
  \bibinfo {author} {\bibfnamefont {Y.-S.}\ \bibnamefont {Lee}}, \bibinfo
  {author} {\bibfnamefont {I.}~\bibnamefont {Souza}}, \bibinfo {author}
  {\bibfnamefont {D.}~\bibnamefont {Vanderbilt}},\ and\ \bibinfo {author}
  {\bibfnamefont {N.}~\bibnamefont {Marzari}},\ }\bibfield  {title} {\bibinfo
  {title} {{Wannier90}: A tool for obtaining maximally-localised {Wannier}
  functions},\ }\href
  {https://doi.org/https://doi.org/10.1016/j.cpc.2007.11.016} {\bibfield
  {journal} {\bibinfo  {journal} {Comput. Phys. Commun.}\ }\textbf {\bibinfo
  {volume} {178}},\ \bibinfo {pages} {685 } (\bibinfo {year}
  {2008})}\BibitemShut {NoStop}%
\bibitem [{\citenamefont {Marzari}\ \emph {et~al.}(2012)\citenamefont
  {Marzari}, \citenamefont {Mostofi}, \citenamefont {Yates}, \citenamefont
  {Souza},\ and\ \citenamefont {Vanderbilt}}]{Marzari2012}%
  \BibitemOpen
  \bibfield  {author} {\bibinfo {author} {\bibfnamefont {N.}~\bibnamefont
  {Marzari}}, \bibinfo {author} {\bibfnamefont {A.~A.}\ \bibnamefont
  {Mostofi}}, \bibinfo {author} {\bibfnamefont {J.~R.}\ \bibnamefont {Yates}},
  \bibinfo {author} {\bibfnamefont {I.}~\bibnamefont {Souza}},\ and\ \bibinfo
  {author} {\bibfnamefont {D.}~\bibnamefont {Vanderbilt}},\ }\bibfield  {title}
  {\bibinfo {title} {Maximally localized {Wannier}functions: Theory and
  applications},\ }\href {https://doi.org/10.1103/RevModPhys.84.1419}
  {\bibfield  {journal} {\bibinfo  {journal} {Rev. Mod. Phys.}\ }\textbf
  {\bibinfo {volume} {84}},\ \bibinfo {pages} {1419} (\bibinfo {year}
  {2012})}\BibitemShut {NoStop}%
\bibitem [{\citenamefont {Mostofi}\ \emph {et~al.}(2014)\citenamefont
  {Mostofi}, \citenamefont {Yates}, \citenamefont {Pizzi}, \citenamefont {Lee},
  \citenamefont {Souza}, \citenamefont {Vanderbilt},\ and\ \citenamefont
  {Marzari}}]{Mostofi2014}%
  \BibitemOpen
  \bibfield  {author} {\bibinfo {author} {\bibfnamefont {A.~A.}\ \bibnamefont
  {Mostofi}}, \bibinfo {author} {\bibfnamefont {J.~R.}\ \bibnamefont {Yates}},
  \bibinfo {author} {\bibfnamefont {G.}~\bibnamefont {Pizzi}}, \bibinfo
  {author} {\bibfnamefont {Y.-S.}\ \bibnamefont {Lee}}, \bibinfo {author}
  {\bibfnamefont {I.}~\bibnamefont {Souza}}, \bibinfo {author} {\bibfnamefont
  {D.}~\bibnamefont {Vanderbilt}},\ and\ \bibinfo {author} {\bibfnamefont
  {N.}~\bibnamefont {Marzari}},\ }\bibfield  {title} {\bibinfo {title} {An
  updated version of {Wannier90}: A tool for obtaining maximally-localised
  {Wannier} functions},\ }\href
  {https://doi.org/https://doi.org/10.1016/j.cpc.2014.05.003} {\bibfield
  {journal} {\bibinfo  {journal} {Comput. Phys. Commun.}\ }\textbf {\bibinfo
  {volume} {185}},\ \bibinfo {pages} {2309 } (\bibinfo {year}
  {2014})}\BibitemShut {NoStop}%
\bibitem [{\citenamefont {Pizzi}\ \emph {et~al.}(2020)\citenamefont {Pizzi},
  \citenamefont {Vitale}, \citenamefont {Arita}, \citenamefont {Blügel},
  \citenamefont {Freimuth}, \citenamefont {G{\'{e}}ranton}, \citenamefont
  {Gibertini}, \citenamefont {Gresch}, \citenamefont {Johnson}, \citenamefont
  {Koretsune}, \citenamefont {Iba{\~{n}}ez-Azpiroz}, \citenamefont {Lee},
  \citenamefont {Lihm}, \citenamefont {Marchand}, \citenamefont {Marrazzo},
  \citenamefont {Mokrousov}, \citenamefont {Mustafa}, \citenamefont {Nohara},
  \citenamefont {Nomura}, \citenamefont {Paulatto}, \citenamefont
  {Ponc{\'{e}}}, \citenamefont {Ponweiser}, \citenamefont {Qiao}, \citenamefont
  {Thöle}, \citenamefont {Tsirkin}, \citenamefont {Wierzbowska}, \citenamefont
  {Marzari}, \citenamefont {Vanderbilt}, \citenamefont {Souza}, \citenamefont
  {Mostofi},\ and\ \citenamefont {Yates}}]{Pizzi2020}%
  \BibitemOpen
  \bibfield  {author} {\bibinfo {author} {\bibfnamefont {G.}~\bibnamefont
  {Pizzi}}, \bibinfo {author} {\bibfnamefont {V.}~\bibnamefont {Vitale}},
  \bibinfo {author} {\bibfnamefont {R.}~\bibnamefont {Arita}}, \bibinfo
  {author} {\bibfnamefont {S.}~\bibnamefont {Blügel}}, \bibinfo {author}
  {\bibfnamefont {F.}~\bibnamefont {Freimuth}}, \bibinfo {author}
  {\bibfnamefont {G.}~\bibnamefont {G{\'{e}}ranton}}, \bibinfo {author}
  {\bibfnamefont {M.}~\bibnamefont {Gibertini}}, \bibinfo {author}
  {\bibfnamefont {D.}~\bibnamefont {Gresch}}, \bibinfo {author} {\bibfnamefont
  {C.}~\bibnamefont {Johnson}}, \bibinfo {author} {\bibfnamefont
  {T.}~\bibnamefont {Koretsune}}, \bibinfo {author} {\bibfnamefont
  {J.}~\bibnamefont {Iba{\~{n}}ez-Azpiroz}}, \bibinfo {author} {\bibfnamefont
  {H.}~\bibnamefont {Lee}}, \bibinfo {author} {\bibfnamefont {J.-M.}\
  \bibnamefont {Lihm}}, \bibinfo {author} {\bibfnamefont {D.}~\bibnamefont
  {Marchand}}, \bibinfo {author} {\bibfnamefont {A.}~\bibnamefont {Marrazzo}},
  \bibinfo {author} {\bibfnamefont {Y.}~\bibnamefont {Mokrousov}}, \bibinfo
  {author} {\bibfnamefont {J.~I.}\ \bibnamefont {Mustafa}}, \bibinfo {author}
  {\bibfnamefont {Y.}~\bibnamefont {Nohara}}, \bibinfo {author} {\bibfnamefont
  {Y.}~\bibnamefont {Nomura}}, \bibinfo {author} {\bibfnamefont
  {L.}~\bibnamefont {Paulatto}}, \bibinfo {author} {\bibfnamefont
  {S.}~\bibnamefont {Ponc{\'{e}}}}, \bibinfo {author} {\bibfnamefont
  {T.}~\bibnamefont {Ponweiser}}, \bibinfo {author} {\bibfnamefont
  {J.}~\bibnamefont {Qiao}}, \bibinfo {author} {\bibfnamefont {F.}~\bibnamefont
  {Thöle}}, \bibinfo {author} {\bibfnamefont {S.~S.}\ \bibnamefont {Tsirkin}},
  \bibinfo {author} {\bibfnamefont {M.}~\bibnamefont {Wierzbowska}}, \bibinfo
  {author} {\bibfnamefont {N.}~\bibnamefont {Marzari}}, \bibinfo {author}
  {\bibfnamefont {D.}~\bibnamefont {Vanderbilt}}, \bibinfo {author}
  {\bibfnamefont {I.}~\bibnamefont {Souza}}, \bibinfo {author} {\bibfnamefont
  {A.~A.}\ \bibnamefont {Mostofi}},\ and\ \bibinfo {author} {\bibfnamefont
  {J.~R.}\ \bibnamefont {Yates}},\ }\bibfield  {title} {\bibinfo {title}
  {{Wannier90} as a community code: new features and applications},\ }\href
  {https://doi.org/10.1088/1361-648x/ab51ff} {\bibfield  {journal} {\bibinfo
  {journal} {J. Phys.: Condens. Matter}\ }\textbf {\bibinfo {volume} {32}},\
  \bibinfo {pages} {165902} (\bibinfo {year} {2020})}\BibitemShut {NoStop}%
\bibitem [{\citenamefont {Haynes}\ \emph {et~al.}(2016)\citenamefont {Haynes},
  \citenamefont {Lide},\ and\ \citenamefont {Bruno}}]{haynes2016crc}%
  \BibitemOpen
  \bibfield  {author} {\bibinfo {author} {\bibfnamefont {W.~M.}\ \bibnamefont
  {Haynes}}, \bibinfo {author} {\bibfnamefont {D.~R.}\ \bibnamefont {Lide}},\
  and\ \bibinfo {author} {\bibfnamefont {T.~J.}\ \bibnamefont {Bruno}},\
  }\href@noop {} {\emph {\bibinfo {title} {CRC handbook of chemistry and
  physics}}}\ (\bibinfo  {publisher} {CRC press},\ \bibinfo {address} {New
  York},\ \bibinfo {year} {2016})\BibitemShut {NoStop}%
\bibitem [{NIS( III)}]{NIST}%
  \BibitemOpen
  \href {http://cccbdb.nist.gov/} {\emph {\bibinfo {title} {{NIST}
  Computational Chemistry Comparison and Benchmark Database}}},\ \bibinfo
  {number} {{NIST} Standard Reference Database Number 101}\ (\bibinfo {year}
  {Release 21, August 2020, edited by R. D. Johnson III})\BibitemShut {NoStop}%
\bibitem [{\citenamefont {Nakamoto}(2006)}]{Nakamoto2006}%
  \BibitemOpen
  \bibfield  {author} {\bibinfo {author} {\bibfnamefont {K.}~\bibnamefont
  {Nakamoto}},\ }\href
  {https://doi.org/https://doi.org/10.1002/0470027320.s4104} {\emph {\bibinfo
  {title} {Handbook of Vibrational Spectroscopy}}}\ (\bibinfo  {publisher}
  {John Wiley \& Sons, Ltd},\ \bibinfo {year} {2006})\BibitemShut {NoStop}%
\bibitem [{\citenamefont {Bradshaw}\ and\ \citenamefont
  {Hoffmann}(1978)}]{BRADSHAW1978513}%
  \BibitemOpen
  \bibfield  {author} {\bibinfo {author} {\bibfnamefont {A.}~\bibnamefont
  {Bradshaw}}\ and\ \bibinfo {author} {\bibfnamefont {F.}~\bibnamefont
  {Hoffmann}},\ }\bibfield  {title} {\bibinfo {title} {The chemisorption of
  carbon monoxide on palladium single crystal surfaces: Ir spectroscopic
  evidence for localised site adsorption},\ }\href
  {https://doi.org/https://doi.org/10.1016/0039-6028(78)90367-9} {\bibfield
  {journal} {\bibinfo  {journal} {Surf. Sci.}\ }\textbf {\bibinfo {volume}
  {72}},\ \bibinfo {pages} {513 } (\bibinfo {year} {1978})}\BibitemShut
  {NoStop}%
\bibitem [{\citenamefont {Gießel}\ \emph {et~al.}(1998)\citenamefont
  {Gießel}, \citenamefont {Schaff}, \citenamefont {Hirschmugl}, \citenamefont
  {Fernandez}, \citenamefont {Schindler}, \citenamefont {Theobald},
  \citenamefont {Bao}, \citenamefont {Lindsay}, \citenamefont {Berndt},
  \citenamefont {Bradshaw}, \citenamefont {Baddeley}, \citenamefont {Lee},
  \citenamefont {Lambert},\ and\ \citenamefont {Woodruff}}]{Giebel1998}%
  \BibitemOpen
  \bibfield  {author} {\bibinfo {author} {\bibfnamefont {T.}~\bibnamefont
  {Gießel}}, \bibinfo {author} {\bibfnamefont {O.}~\bibnamefont {Schaff}},
  \bibinfo {author} {\bibfnamefont {C.}~\bibnamefont {Hirschmugl}}, \bibinfo
  {author} {\bibfnamefont {V.}~\bibnamefont {Fernandez}}, \bibinfo {author}
  {\bibfnamefont {K.-M.}\ \bibnamefont {Schindler}}, \bibinfo {author}
  {\bibfnamefont {A.}~\bibnamefont {Theobald}}, \bibinfo {author}
  {\bibfnamefont {S.}~\bibnamefont {Bao}}, \bibinfo {author} {\bibfnamefont
  {R.}~\bibnamefont {Lindsay}}, \bibinfo {author} {\bibfnamefont
  {W.}~\bibnamefont {Berndt}}, \bibinfo {author} {\bibfnamefont
  {A.}~\bibnamefont {Bradshaw}}, \bibinfo {author} {\bibfnamefont
  {C.}~\bibnamefont {Baddeley}}, \bibinfo {author} {\bibfnamefont
  {A.}~\bibnamefont {Lee}}, \bibinfo {author} {\bibfnamefont {R.}~\bibnamefont
  {Lambert}},\ and\ \bibinfo {author} {\bibfnamefont {D.}~\bibnamefont
  {Woodruff}},\ }\bibfield  {title} {\bibinfo {title} {A photoelectron
  diffraction study of ordered structures in the chemisorption system
  {Pd}(111)-{CO}},\ }\href
  {https://doi.org/https://doi.org/10.1016/S0039-6028(98)00098-3} {\bibfield
  {journal} {\bibinfo  {journal} {Surf. Sci.}\ }\textbf {\bibinfo {volume}
  {406}},\ \bibinfo {pages} {90 } (\bibinfo {year} {1998})}\BibitemShut
  {NoStop}%
\bibitem [{\citenamefont {Anisimov}\ \emph {et~al.}(1974)\citenamefont
  {Anisimov}, \citenamefont {Kapeliovich},\ and\ \citenamefont
  {Perel'Man}}]{Anisimov1974}%
  \BibitemOpen
  \bibfield  {author} {\bibinfo {author} {\bibfnamefont {S.~I.}\ \bibnamefont
  {Anisimov}}, \bibinfo {author} {\bibfnamefont {B.~L.}\ \bibnamefont
  {Kapeliovich}},\ and\ \bibinfo {author} {\bibfnamefont {T.~L.}\ \bibnamefont
  {Perel'Man}},\ }\bibfield  {title} {\bibinfo {title} {Electron emission from
  metal surfaces exposed to ultrashort laser pulses},\ }\href@noop {}
  {\bibfield  {journal} {\bibinfo  {journal} {Sov. Phys. JETP}\ }\textbf
  {\bibinfo {volume} {39}},\ \bibinfo {pages} {375} (\bibinfo {year}
  {1974})}\BibitemShut {NoStop}%
\bibitem [{\citenamefont {Ashcroft}\ and\ \citenamefont
  {Mermin}(1988)}]{Ashcroft1988}%
  \BibitemOpen
  \bibfield  {author} {\bibinfo {author} {\bibfnamefont {N.~W.}\ \bibnamefont
  {Ashcroft}}\ and\ \bibinfo {author} {\bibfnamefont {N.~D.}\ \bibnamefont
  {Mermin}},\ }\href@noop {} {\emph {\bibinfo {title} {Solid state physics}}},\
  edited by\ \bibinfo {editor} {\bibfnamefont {P.}~\bibnamefont
  {Saunders~College}}\ (\bibinfo {year} {1988})\BibitemShut {NoStop}%
\bibitem [{\citenamefont {Kittel}(1986)}]{Kittel1986}%
  \BibitemOpen
  \bibfield  {author} {\bibinfo {author} {\bibfnamefont {C.}~\bibnamefont
  {Kittel}},\ }\href@noop {} {\emph {\bibinfo {title} {Introduction to solid
  state physics}}},\ \bibinfo {edition} {6th}\ ed.,\ edited by\ \bibinfo
  {editor} {\bibfnamefont {N.~Y.}\ \bibnamefont {Wiley \&~Sons}}\ (\bibinfo
  {year} {1986})\BibitemShut {NoStop}%
\bibitem [{\citenamefont {Petrov}\ \emph {et~al.}(2013)\citenamefont {Petrov},
  \citenamefont {Inogamov},\ and\ \citenamefont {Migdal}}]{Petrov2013}%
  \BibitemOpen
  \bibfield  {author} {\bibinfo {author} {\bibfnamefont {Y.~V.}\ \bibnamefont
  {Petrov}}, \bibinfo {author} {\bibfnamefont {N.~A.}\ \bibnamefont
  {Inogamov}},\ and\ \bibinfo {author} {\bibfnamefont {K.~P.}\ \bibnamefont
  {Migdal}},\ }\bibfield  {title} {\bibinfo {title} {Thermal conductivity and
  the electron-ion heat transfer coefficient in condensed media with a strongly
  excited electron subsystem},\ }\href
  {https://doi.org/10.1134/S0021364013010098} {\bibfield  {journal} {\bibinfo
  {journal} {JETP Letters}\ }\textbf {\bibinfo {volume} {97}},\ \bibinfo
  {pages} {20} (\bibinfo {year} {2013})}\BibitemShut {NoStop}%
\bibitem [{\citenamefont {Johnson}\ and\ \citenamefont
  {Christy}()}]{Johnson1974}%
  \BibitemOpen
  \bibfield  {author} {\bibinfo {author} {\bibfnamefont {P.}~\bibnamefont
  {Johnson}}\ and\ \bibinfo {author} {\bibfnamefont {R.}~\bibnamefont
  {Christy}},\ }\bibfield  {title} {\bibinfo {title} {Optical constants of
  transition metals: {Ti, V, Cr, Mn, Fe, Co, Ni, and Pd}},\ }\href
  {https://doi.org/10.1103/PhysRevB.9.5056} {\bibfield  {journal} {\bibinfo
  {journal} {Phys. Rev. B}\ }\textbf {\bibinfo {volume} {9}},\ \bibinfo {pages}
  {5056}}\BibitemShut {NoStop}%
\bibitem [{\citenamefont {Szymanski}\ \emph {et~al.}(2007)\citenamefont
  {Szymanski}, \citenamefont {Harris},\ and\ \citenamefont
  {Camillone}}]{Szymanski2007}%
  \BibitemOpen
  \bibfield  {author} {\bibinfo {author} {\bibfnamefont {P.}~\bibnamefont
  {Szymanski}}, \bibinfo {author} {\bibfnamefont {A.}~\bibnamefont {Harris}},\
  and\ \bibinfo {author} {\bibfnamefont {N.}~\bibnamefont {Camillone}},\
  }\bibfield  {title} {\bibinfo {title} {Temperature-dependent
  electron-mediated coupling in subpicosecond photoinduced desorption},\ }\href
  {https://doi.org/https://doi.org/10.1016/j.susc.2007.06.004} {\bibfield
  {journal} {\bibinfo  {journal} {Surf. Sci.}\ }\textbf {\bibinfo {volume}
  {601}},\ \bibinfo {pages} {3335} (\bibinfo {year} {2007})}\BibitemShut
  {NoStop}%
\bibitem [{\citenamefont {Hong}\ \emph {et~al.}(2016)\citenamefont {Hong},
  \citenamefont {Xu}, \citenamefont {Camillone}, \citenamefont {White},\ and\
  \citenamefont {Camillone}}]{Hong2016}%
  \BibitemOpen
  \bibfield  {author} {\bibinfo {author} {\bibfnamefont {S.-Y.}\ \bibnamefont
  {Hong}}, \bibinfo {author} {\bibfnamefont {P.}~\bibnamefont {Xu}}, \bibinfo
  {author} {\bibfnamefont {N.~R.}\ \bibnamefont {Camillone}}, \bibinfo {author}
  {\bibfnamefont {M.~G.}\ \bibnamefont {White}},\ and\ \bibinfo {author}
  {\bibfnamefont {N.}~\bibnamefont {Camillone}},\ }\bibfield  {title} {\bibinfo
  {title} {Adlayer structure dependent ultrafast desorption dynamics in carbon
  monoxide adsorbed on {Pd} (111)},\ }\href {https://doi.org/10.1063/1.4954408}
  {\bibfield  {journal} {\bibinfo  {journal} {J. Chem. Phys.}\ }\textbf
  {\bibinfo {volume} {145}},\ \bibinfo {pages} {014704} (\bibinfo {year}
  {2016})}\BibitemShut {NoStop}%
\bibitem [{\citenamefont {Li}\ and\ \citenamefont {Ji}(2022)}]{Li2022}%
  \BibitemOpen
  \bibfield  {author} {\bibinfo {author} {\bibfnamefont {Y.}~\bibnamefont
  {Li}}\ and\ \bibinfo {author} {\bibfnamefont {P.}~\bibnamefont {Ji}},\
  }\bibfield  {title} {\bibinfo {title} {Ab initio calculation of electron
  temperature dependent electron heat capacity and electron-phonon coupling
  factor of noble metals},\ }\href
  {https://doi.org/https://doi.org/10.1016/j.commatsci.2021.110959} {\bibfield
  {journal} {\bibinfo  {journal} {Comp. Mater Sci.}\ }\textbf {\bibinfo
  {volume} {202}},\ \bibinfo {pages} {110959} (\bibinfo {year}
  {2022})}\BibitemShut {NoStop}%
\bibitem [{\citenamefont {Luttinger}(1960)}]{luttinger60}%
  \BibitemOpen
  \bibfield  {author} {\bibinfo {author} {\bibfnamefont {J.~M.}\ \bibnamefont
  {Luttinger}},\ }\bibfield  {title} {\bibinfo {title} {Fermi surface and some
  simple equilibrium properties of a system of interacting fermions},\ }\href
  {https://doi.org/10.1103/PhysRev.119.1153} {\bibfield  {journal} {\bibinfo
  {journal} {Phys. Rev.}\ }\textbf {\bibinfo {volume} {119}},\ \bibinfo {pages}
  {1153} (\bibinfo {year} {1960})}\BibitemShut {NoStop}%
\bibitem [{\citenamefont {Andreatta}\ \emph {et~al.}(2019)\citenamefont
  {Andreatta}, \citenamefont {Rostami}, \citenamefont {\ifmmode~\check{C}\else
  \v{C}\fi{}abo}, \citenamefont {Bianchi}, \citenamefont {Sanders},
  \citenamefont {Biswas}, \citenamefont {Cacho}, \citenamefont {Jones},
  \citenamefont {Chapman}, \citenamefont {Springate}, \citenamefont {King},
  \citenamefont {Miwa}, \citenamefont {Balatsky}, \citenamefont {Ulstrup},\
  and\ \citenamefont {Hofmann}}]{andreatta19}%
  \BibitemOpen
  \bibfield  {author} {\bibinfo {author} {\bibfnamefont {F.}~\bibnamefont
  {Andreatta}}, \bibinfo {author} {\bibfnamefont {H.}~\bibnamefont {Rostami}},
  \bibinfo {author} {\bibfnamefont {A.~G.}\ \bibnamefont
  {\ifmmode~\check{C}\else \v{C}\fi{}abo}}, \bibinfo {author} {\bibfnamefont
  {M.}~\bibnamefont {Bianchi}}, \bibinfo {author} {\bibfnamefont {C.~E.}\
  \bibnamefont {Sanders}}, \bibinfo {author} {\bibfnamefont {D.}~\bibnamefont
  {Biswas}}, \bibinfo {author} {\bibfnamefont {C.}~\bibnamefont {Cacho}},
  \bibinfo {author} {\bibfnamefont {A.~J.~H.}\ \bibnamefont {Jones}}, \bibinfo
  {author} {\bibfnamefont {R.~T.}\ \bibnamefont {Chapman}}, \bibinfo {author}
  {\bibfnamefont {E.}~\bibnamefont {Springate}}, \bibinfo {author}
  {\bibfnamefont {P.~D.~C.}\ \bibnamefont {King}}, \bibinfo {author}
  {\bibfnamefont {J.~A.}\ \bibnamefont {Miwa}}, \bibinfo {author}
  {\bibfnamefont {A.}~\bibnamefont {Balatsky}}, \bibinfo {author}
  {\bibfnamefont {S.}~\bibnamefont {Ulstrup}},\ and\ \bibinfo {author}
  {\bibfnamefont {P.}~\bibnamefont {Hofmann}},\ }\bibfield  {title} {\bibinfo
  {title} {Transient hot electron dynamics in single-layer
  {${\mathrm{TaS}}_{2}$}},\ }\href {https://doi.org/10.1103/PhysRevB.99.165421}
  {\bibfield  {journal} {\bibinfo  {journal} {Phys. Rev. B}\ }\textbf {\bibinfo
  {volume} {99}},\ \bibinfo {pages} {165421} (\bibinfo {year}
  {2019})}\BibitemShut {NoStop}%
\end{thebibliography}%


\begin{thebibliography}{57}%
\makeatletter
\providecommand \@ifxundefined [1]{%
 \@ifx{#1\undefined}
}%
\providecommand \@ifnum [1]{%
 \ifnum #1\expandafter \@firstoftwo
 \else \expandafter \@secondoftwo
 \fi
}%
\providecommand \@ifx [1]{%
 \ifx #1\expandafter \@firstoftwo
 \else \expandafter \@secondoftwo
 \fi
}%
\providecommand \natexlab [1]{#1}%
\providecommand \enquote  [1]{``#1''}%
\providecommand \bibnamefont  [1]{#1}%
\providecommand \bibfnamefont [1]{#1}%
\providecommand \citenamefont [1]{#1}%
\providecommand \href@noop [0]{\@secondoftwo}%
\providecommand \href [0]{\begingroup \@sanitize@url \@href}%
\providecommand \@href[1]{\@@startlink{#1}\@@href}%
\providecommand \@@href[1]{\endgroup#1\@@endlink}%
\providecommand \@sanitize@url [0]{\catcode `\\12\catcode `\$12\catcode
  `\&12\catcode `\#12\catcode `\^12\catcode `\_12\catcode `\%12\relax}%
\providecommand \@@startlink[1]{}%
\providecommand \@@endlink[0]{}%
\providecommand \url  [0]{\begingroup\@sanitize@url \@url }%
\providecommand \@url [1]{\endgroup\@href {#1}{\urlprefix }}%
\providecommand \urlprefix  [0]{URL }%
\providecommand \Eprint [0]{\href }%
\providecommand \doibase [0]{https://doi.org/}%
\providecommand \selectlanguage [0]{\@gobble}%
\providecommand \bibinfo  [0]{\@secondoftwo}%
\providecommand \bibfield  [0]{\@secondoftwo}%
\providecommand \translation [1]{[#1]}%
\providecommand \BibitemOpen [0]{}%
\providecommand \bibitemStop [0]{}%
\providecommand \bibitemNoStop [0]{.\EOS\space}%
\providecommand \EOS [0]{\spacefactor3000\relax}%
\providecommand \BibitemShut  [1]{\csname bibitem#1\endcsname}%
\let\auto@bib@innerbib\@empty
\bibitem [{\citenamefont {Budde}\ \emph {et~al.}(1991)\citenamefont {Budde},
  \citenamefont {Heinz}, \citenamefont {Loy}, \citenamefont {Misewich},
  \citenamefont {de~Rougemont},\ and\ \citenamefont {Zacharias}}]{budde91}%
  \BibitemOpen
  \bibfield  {author} {\bibinfo {author} {\bibfnamefont {F.}~\bibnamefont
  {Budde}}, \bibinfo {author} {\bibfnamefont {T.~F.}\ \bibnamefont {Heinz}},
  \bibinfo {author} {\bibfnamefont {M.~M.~T.}\ \bibnamefont {Loy}}, \bibinfo
  {author} {\bibfnamefont {J.~A.}\ \bibnamefont {Misewich}}, \bibinfo {author}
  {\bibfnamefont {F.}~\bibnamefont {de~Rougemont}},\ and\ \bibinfo {author}
  {\bibfnamefont {H.}~\bibnamefont {Zacharias}},\ }\bibfield  {title} {\bibinfo
  {title} {Femtosecond time-resolved measurement of desorption},\ }\href
  {https://doi.org/10.1103/PhysRevLett.66.3024} {\bibfield  {journal} {\bibinfo
   {journal} {Phys. Rev. Lett.}\ }\textbf {\bibinfo {volume} {66}},\ \bibinfo
  {pages} {3024} (\bibinfo {year} {1991})}\BibitemShut {NoStop}%
\bibitem [{\citenamefont {Prybyla}\ \emph {et~al.}(1992)\citenamefont
  {Prybyla}, \citenamefont {Tom},\ and\ \citenamefont
  {Aumiller}}]{Prybyla1992}%
  \BibitemOpen
  \bibfield  {author} {\bibinfo {author} {\bibfnamefont {J.~A.}\ \bibnamefont
  {Prybyla}}, \bibinfo {author} {\bibfnamefont {H.~W.~K.}\ \bibnamefont
  {Tom}},\ and\ \bibinfo {author} {\bibfnamefont {G.~D.}\ \bibnamefont
  {Aumiller}},\ }\bibfield  {title} {\bibinfo {title} {Femtosecond
  time-resolved surface reaction: Desorption of {CO} from {Cu}(111) in $<$ 325
  fsec},\ }\href {https://doi.org/10.1103/PhysRevLett.68.503} {\bibfield
  {journal} {\bibinfo  {journal} {Phys. Rev. Lett.}\ }\textbf {\bibinfo
  {volume} {68}},\ \bibinfo {pages} {503} (\bibinfo {year} {1992})}\BibitemShut
  {NoStop}%
\bibitem [{\citenamefont {Lawrenz}\ \emph {et~al.}(2009)\citenamefont
  {Lawrenz}, \citenamefont {St\'ep\'an}, \citenamefont {G\"udde},\ and\
  \citenamefont {H\"ofer}}]{Lawrenz2009}%
  \BibitemOpen
  \bibfield  {author} {\bibinfo {author} {\bibfnamefont {M.}~\bibnamefont
  {Lawrenz}}, \bibinfo {author} {\bibfnamefont {K.}~\bibnamefont {St\'ep\'an}},
  \bibinfo {author} {\bibfnamefont {J.}~\bibnamefont {G\"udde}},\ and\ \bibinfo
  {author} {\bibfnamefont {U.}~\bibnamefont {H\"ofer}},\ }\bibfield  {title}
  {\bibinfo {title} {Time-domain investigation of laser-induced diffusion of
  {CO} on a vicinal {Pt}(111) surface},\ }\href
  {https://doi.org/10.1103/PhysRevB.80.075429} {\bibfield  {journal} {\bibinfo
  {journal} {Phys. Rev. B}\ }\textbf {\bibinfo {volume} {80}},\ \bibinfo
  {pages} {075429} (\bibinfo {year} {2009})}\BibitemShut {NoStop}%
\bibitem [{\citenamefont {Frischkorn}\ and\ \citenamefont
  {Wolf}(2006)}]{frischkorncr06}%
  \BibitemOpen
  \bibfield  {author} {\bibinfo {author} {\bibfnamefont {C.}~\bibnamefont
  {Frischkorn}}\ and\ \bibinfo {author} {\bibfnamefont {M.}~\bibnamefont
  {Wolf}},\ }\bibfield  {title} {\bibinfo {title} {Femtochemistry at metal
  surfaces: Nonadiabatic reaction dynamics},\ }\href
  {https://doi.org/10.1021/cr050161r} {\bibfield  {journal} {\bibinfo
  {journal} {Chem. Rev.}\ }\textbf {\bibinfo {volume} {106}},\ \bibinfo {pages}
  {4207} (\bibinfo {year} {2006})},\ \Eprint
  {https://arxiv.org/abs/http://dx.doi.org/10.1021/cr050161r}
  {http://dx.doi.org/10.1021/cr050161r} \BibitemShut {NoStop}%
\bibitem [{\citenamefont {Park}\ \emph {et~al.}(2015)\citenamefont {Park},
  \citenamefont {Kim}, \citenamefont {Lee},\ and\ \citenamefont
  {Nedrygailov}}]{Park2015}%
  \BibitemOpen
  \bibfield  {author} {\bibinfo {author} {\bibfnamefont {J.~Y.}\ \bibnamefont
  {Park}}, \bibinfo {author} {\bibfnamefont {S.~M.}\ \bibnamefont {Kim}},
  \bibinfo {author} {\bibfnamefont {H.}~\bibnamefont {Lee}},\ and\ \bibinfo
  {author} {\bibfnamefont {I.~I.}\ \bibnamefont {Nedrygailov}},\ }\bibfield
  {title} {\bibinfo {title} {Hot-electron-mediated surface chemistry: Toward
  electronic control of catalytic activity},\ }\href
  {https://doi.org/10.1021/acs.accounts.5b00170} {\bibfield  {journal}
  {\bibinfo  {journal} {Acc. Chem. Res}\ }\textbf {\bibinfo {volume} {48}},\
  \bibinfo {pages} {2475} (\bibinfo {year} {2015})},\ \bibinfo {note} {pMID:
  26181684},\ \Eprint
  {https://arxiv.org/abs/https://doi.org/10.1021/acs.accounts.5b00170}
  {https://doi.org/10.1021/acs.accounts.5b00170} \BibitemShut {NoStop}%
\bibitem [{\citenamefont {Arnolds}\ and\ \citenamefont
  {Bonn}(2010)}]{arnolds2010}%
  \BibitemOpen
  \bibfield  {author} {\bibinfo {author} {\bibfnamefont {H.}~\bibnamefont
  {Arnolds}}\ and\ \bibinfo {author} {\bibfnamefont {M.}~\bibnamefont {Bonn}},\
  }\bibfield  {title} {\bibinfo {title} {Ultrafast surface vibrational
  dynamics},\ }\href
  {https://doi.org/https://doi.org/10.1016/j.surfrep.2009.12.001} {\bibfield
  {journal} {\bibinfo  {journal} {Surf. Sci. Rep.}\ }\textbf {\bibinfo {volume}
  {65}},\ \bibinfo {pages} {45} (\bibinfo {year} {2010})}\BibitemShut {NoStop}%
\bibitem [{\citenamefont {Yampolsky}\ \emph {et~al.}(2014)\citenamefont
  {Yampolsky}, \citenamefont {Fishman}, \citenamefont {Dey}, \citenamefont
  {Hulkko}, \citenamefont {Banik}, \citenamefont {Potma},\ and\ \citenamefont
  {Apkarian}}]{Yampolsky2014}%
  \BibitemOpen
  \bibfield  {author} {\bibinfo {author} {\bibfnamefont {S.}~\bibnamefont
  {Yampolsky}}, \bibinfo {author} {\bibfnamefont {D.~A.}\ \bibnamefont
  {Fishman}}, \bibinfo {author} {\bibfnamefont {S.}~\bibnamefont {Dey}},
  \bibinfo {author} {\bibfnamefont {E.}~\bibnamefont {Hulkko}}, \bibinfo
  {author} {\bibfnamefont {M.}~\bibnamefont {Banik}}, \bibinfo {author}
  {\bibfnamefont {E.~O.}\ \bibnamefont {Potma}},\ and\ \bibinfo {author}
  {\bibfnamefont {V.~A.}\ \bibnamefont {Apkarian}},\ }\bibfield  {title}
  {\bibinfo {title} {Seeing a single molecule vibrate through time-resolved
  coherent anti-{Stokes} {Raman} scattering},\ }\href
  {https://doi.org/10.1038/nphoton.2014.143} {\bibfield  {journal} {\bibinfo
  {journal} {Nat. Phot.}\ }\textbf {\bibinfo {volume} {8}},\ \bibinfo {pages}
  {650} (\bibinfo {year} {2014})}\BibitemShut {NoStop}%
\bibitem [{\citenamefont {Omiya}\ and\ \citenamefont
  {Arnolds}(2014)}]{Omiya2014}%
  \BibitemOpen
  \bibfield  {author} {\bibinfo {author} {\bibfnamefont {T.}~\bibnamefont
  {Omiya}}\ and\ \bibinfo {author} {\bibfnamefont {H.}~\bibnamefont
  {Arnolds}},\ }\bibfield  {title} {\bibinfo {title} {Coverage dependent
  non-adiabaticity of {CO} on a copper surface},\ }\href
  {https://doi.org/10.1063/1.4902540} {\bibfield  {journal} {\bibinfo
  {journal} {J. Chem. Phys.}\ }\textbf {\bibinfo {volume} {141}},\ \bibinfo
  {pages} {214705} (\bibinfo {year} {2014})}\BibitemShut {NoStop}%
\bibitem [{\citenamefont {Bonn}\ \emph {et~al.}(2000)\citenamefont {Bonn},
  \citenamefont {Hess}, \citenamefont {Funk}, \citenamefont {Miners},
  \citenamefont {Persson}, \citenamefont {Wolf},\ and\ \citenamefont
  {Ertl}}]{Bonn2000}%
  \BibitemOpen
  \bibfield  {author} {\bibinfo {author} {\bibfnamefont {M.}~\bibnamefont
  {Bonn}}, \bibinfo {author} {\bibfnamefont {C.}~\bibnamefont {Hess}}, \bibinfo
  {author} {\bibfnamefont {S.}~\bibnamefont {Funk}}, \bibinfo {author}
  {\bibfnamefont {J.~H.}\ \bibnamefont {Miners}}, \bibinfo {author}
  {\bibfnamefont {B.~N.~J.}\ \bibnamefont {Persson}}, \bibinfo {author}
  {\bibfnamefont {M.}~\bibnamefont {Wolf}},\ and\ \bibinfo {author}
  {\bibfnamefont {G.}~\bibnamefont {Ertl}},\ }\bibfield  {title} {\bibinfo
  {title} {Femtosecond surface vibrational spectroscopy of {CO} adsorbed on
  {Ru}(001) during desorption},\ }\href
  {https://doi.org/10.1103/PhysRevLett.84.4653} {\bibfield  {journal} {\bibinfo
   {journal} {Phys. Rev. Lett.}\ }\textbf {\bibinfo {volume} {84}},\ \bibinfo
  {pages} {4653} (\bibinfo {year} {2000})}\BibitemShut {NoStop}%
\bibitem [{\citenamefont {Lane}\ \emph {et~al.}(2006)\citenamefont {Lane},
  \citenamefont {King}, \citenamefont {Liu},\ and\ \citenamefont
  {Arnolds}}]{Lane2006}%
  \BibitemOpen
  \bibfield  {author} {\bibinfo {author} {\bibfnamefont {I.~M.}\ \bibnamefont
  {Lane}}, \bibinfo {author} {\bibfnamefont {D.~A.}\ \bibnamefont {King}},
  \bibinfo {author} {\bibfnamefont {Z.-P.}\ \bibnamefont {Liu}},\ and\ \bibinfo
  {author} {\bibfnamefont {H.}~\bibnamefont {Arnolds}},\ }\bibfield  {title}
  {\bibinfo {title} {Real-time observation of nonadiabatic surface dynamics:
  The first picosecond in the dissociation of {NO} on iridium},\ }\href
  {https://doi.org/10.1103/PhysRevLett.97.186105} {\bibfield  {journal}
  {\bibinfo  {journal} {Phys. Rev. Lett.}\ }\textbf {\bibinfo {volume} {97}},\
  \bibinfo {pages} {186105} (\bibinfo {year} {2006})}\BibitemShut {NoStop}%
\bibitem [{\citenamefont {Lane}\ \emph {et~al.}(2007)\citenamefont {Lane},
  \citenamefont {Liu}, \citenamefont {King},\ and\ \citenamefont
  {Arnolds}}]{Lane2007}%
  \BibitemOpen
  \bibfield  {author} {\bibinfo {author} {\bibfnamefont {I.~M.}\ \bibnamefont
  {Lane}}, \bibinfo {author} {\bibfnamefont {Z.-P.}\ \bibnamefont {Liu}},
  \bibinfo {author} {\bibfnamefont {D.~A.}\ \bibnamefont {King}},\ and\
  \bibinfo {author} {\bibfnamefont {H.}~\bibnamefont {Arnolds}},\ }\bibfield
  {title} {\bibinfo {title} {Ultrafast vibrational dynamics of {NO} and {CO}
  adsorbed on an iridium surface},\ }\href {https://doi.org/10.1021/jp071831v}
  {\bibfield  {journal} {\bibinfo  {journal} {J. Phys. Chem. C}\ }\textbf
  {\bibinfo {volume} {111}},\ \bibinfo {pages} {14198} (\bibinfo {year}
  {2007})}\BibitemShut {NoStop}%
\bibitem [{\citenamefont {Fournier}\ \emph {et~al.}(2004)\citenamefont
  {Fournier}, \citenamefont {Zheng}, \citenamefont {Carrez}, \citenamefont
  {Dubost},\ and\ \citenamefont {Bourguignon}}]{Fournier2004}%
  \BibitemOpen
  \bibfield  {author} {\bibinfo {author} {\bibfnamefont {F.}~\bibnamefont
  {Fournier}}, \bibinfo {author} {\bibfnamefont {W.}~\bibnamefont {Zheng}},
  \bibinfo {author} {\bibfnamefont {S.}~\bibnamefont {Carrez}}, \bibinfo
  {author} {\bibfnamefont {H.}~\bibnamefont {Dubost}},\ and\ \bibinfo {author}
  {\bibfnamefont {B.}~\bibnamefont {Bourguignon}},\ }\bibfield  {title}
  {\bibinfo {title} {Ultrafast laser excitation of
  $\mathrm{CO}/\mathrm{Pt}(111)$ probed by sum frequency generation: Coverage
  dependent desorption efficiency},\ }\href
  {https://doi.org/10.1103/PhysRevLett.92.216102} {\bibfield  {journal}
  {\bibinfo  {journal} {Phys. Rev. Lett.}\ }\textbf {\bibinfo {volume} {92}},\
  \bibinfo {pages} {216102} (\bibinfo {year} {2004})}\BibitemShut {NoStop}%
\bibitem [{\citenamefont {Watanabe}\ \emph {et~al.}(2010)\citenamefont
  {Watanabe}, \citenamefont {Inoue}, \citenamefont {Nakai},\ and\ \citenamefont
  {Matsumoto}}]{Watanabe2010}%
  \BibitemOpen
  \bibfield  {author} {\bibinfo {author} {\bibfnamefont {K.}~\bibnamefont
  {Watanabe}}, \bibinfo {author} {\bibfnamefont {K.-i.}\ \bibnamefont {Inoue}},
  \bibinfo {author} {\bibfnamefont {I.~F.}\ \bibnamefont {Nakai}},\ and\
  \bibinfo {author} {\bibfnamefont {Y.}~\bibnamefont {Matsumoto}},\ }\bibfield
  {title} {\bibinfo {title} {Nonadiabatic coupling between {C-O} stretching and
  {Pt} substrate electrons enhanced by frustrated mode excitations},\ }\href
  {https://doi.org/10.1103/PhysRevB.81.241408} {\bibfield  {journal} {\bibinfo
  {journal} {Phys. Rev. B}\ }\textbf {\bibinfo {volume} {81}},\ \bibinfo
  {pages} {241408(R)} (\bibinfo {year} {2010})}\BibitemShut {NoStop}%
\bibitem [{\citenamefont {Inoue}\ \emph {et~al.}(2012)\citenamefont {Inoue},
  \citenamefont {Watanabe},\ and\ \citenamefont {Matsumoto}}]{Inoue2012}%
  \BibitemOpen
  \bibfield  {author} {\bibinfo {author} {\bibfnamefont {K.-i.}\ \bibnamefont
  {Inoue}}, \bibinfo {author} {\bibfnamefont {K.}~\bibnamefont {Watanabe}},\
  and\ \bibinfo {author} {\bibfnamefont {Y.}~\bibnamefont {Matsumoto}},\
  }\bibfield  {title} {\bibinfo {title} {Instantaneous vibrational frequencies
  of diffusing and desorbing adsorbates: {CO}/{Pt}(111)},\ }\href
  {https://doi.org/10.1063/1.4733720} {\bibfield  {journal} {\bibinfo
  {journal} {J. Chem. Phys.}\ }\textbf {\bibinfo {volume} {137}},\ \bibinfo
  {pages} {024704} (\bibinfo {year} {2012})},\ \Eprint
  {https://arxiv.org/abs/https://doi.org/10.1063/1.4733720}
  {https://doi.org/10.1063/1.4733720} \BibitemShut {NoStop}%
\bibitem [{\citenamefont {Inoue}\ \emph {et~al.}(2016)\citenamefont {Inoue},
  \citenamefont {Watanabe}, \citenamefont {Sugimoto}, \citenamefont
  {Matsumoto},\ and\ \citenamefont {Yasuike}}]{Inoue2016}%
  \BibitemOpen
  \bibfield  {author} {\bibinfo {author} {\bibfnamefont {K.-i.}\ \bibnamefont
  {Inoue}}, \bibinfo {author} {\bibfnamefont {K.}~\bibnamefont {Watanabe}},
  \bibinfo {author} {\bibfnamefont {T.}~\bibnamefont {Sugimoto}}, \bibinfo
  {author} {\bibfnamefont {Y.}~\bibnamefont {Matsumoto}},\ and\ \bibinfo
  {author} {\bibfnamefont {T.}~\bibnamefont {Yasuike}},\ }\bibfield  {title}
  {\bibinfo {title} {Disentangling multidimensional nonequilibrium dynamics of
  adsorbates: {CO} desorption from {Cu}(100)},\ }\href
  {https://doi.org/10.1103/PhysRevLett.117.186101} {\bibfield  {journal}
  {\bibinfo  {journal} {Phys. Rev. Lett.}\ }\textbf {\bibinfo {volume} {117}},\
  \bibinfo {pages} {186101} (\bibinfo {year} {2016})}\BibitemShut {NoStop}%
\bibitem [{\citenamefont {Omiya}\ \emph {et~al.}(2019)\citenamefont {Omiya},
  \citenamefont {Kim}, \citenamefont {Raval},\ and\ \citenamefont
  {Arnolds}}]{Omiya2019}%
  \BibitemOpen
  \bibfield  {author} {\bibinfo {author} {\bibfnamefont {T.}~\bibnamefont
  {Omiya}}, \bibinfo {author} {\bibfnamefont {Y.}~\bibnamefont {Kim}}, \bibinfo
  {author} {\bibfnamefont {R.}~\bibnamefont {Raval}},\ and\ \bibinfo {author}
  {\bibfnamefont {H.}~\bibnamefont {Arnolds}},\ }\bibfield  {title} {\bibinfo
  {title} {Ultrafast vibrational dynamics of {CO} ligands on {RuTPP/Cu(110)}
  under photodesorption conditions},\ }\href
  {https://www.mdpi.com/2571-9637/2/1/10} {\bibfield  {journal} {\bibinfo
  {journal} {Surfaces}\ }\textbf {\bibinfo {volume} {2}},\ \bibinfo {pages}
  {117} (\bibinfo {year} {2019})}\BibitemShut {NoStop}%
\bibitem [{\citenamefont {Novko}\ \emph {et~al.}(2019)\citenamefont {Novko},
  \citenamefont {Tremblay}, \citenamefont {Alducin},\ and\ \citenamefont
  {Juaristi}}]{Novko2019}%
  \BibitemOpen
  \bibfield  {author} {\bibinfo {author} {\bibfnamefont {D.}~\bibnamefont
  {Novko}}, \bibinfo {author} {\bibfnamefont {J.~C.}\ \bibnamefont {Tremblay}},
  \bibinfo {author} {\bibfnamefont {M.}~\bibnamefont {Alducin}},\ and\ \bibinfo
  {author} {\bibfnamefont {J.~I.}\ \bibnamefont {Juaristi}},\ }\bibfield
  {title} {\bibinfo {title} {Ultrafast transient dynamics of adsorbates on
  surfaces deciphered: The case of {CO} on {Cu}(100)},\ }\href
  {https://doi.org/10.1103/PhysRevLett.122.016806} {\bibfield  {journal}
  {\bibinfo  {journal} {Phys. Rev. Lett.}\ }\textbf {\bibinfo {volume} {122}},\
  \bibinfo {pages} {016806} (\bibinfo {year} {2019})}\BibitemShut {NoStop}%
\bibitem [{\citenamefont {Baddorf}\ and\ \citenamefont
  {Plummer}(1991)}]{Baddorf1991}%
  \BibitemOpen
  \bibfield  {author} {\bibinfo {author} {\bibfnamefont {A.~P.}\ \bibnamefont
  {Baddorf}}\ and\ \bibinfo {author} {\bibfnamefont {E.~W.}\ \bibnamefont
  {Plummer}},\ }\bibfield  {title} {\bibinfo {title} {Enhanced surface
  anharmonicity observed in vibrations on {Cu(110)}},\ }\href
  {https://doi.org/10.1103/PhysRevLett.66.2770} {\bibfield  {journal} {\bibinfo
   {journal} {Phys. Rev. Lett.}\ }\textbf {\bibinfo {volume} {66}},\ \bibinfo
  {pages} {2770} (\bibinfo {year} {1991})}\BibitemShut {NoStop}%
\bibitem [{\citenamefont {Bracco}\ \emph {et~al.}(1996)\citenamefont {Bracco},
  \citenamefont {Bruschi}, \citenamefont {Pedemonte},\ and\ \citenamefont
  {Tatarek}}]{Bracco1996}%
  \BibitemOpen
  \bibfield  {author} {\bibinfo {author} {\bibfnamefont {G.}~\bibnamefont
  {Bracco}}, \bibinfo {author} {\bibfnamefont {L.}~\bibnamefont {Bruschi}},
  \bibinfo {author} {\bibfnamefont {L.}~\bibnamefont {Pedemonte}},\ and\
  \bibinfo {author} {\bibfnamefont {R.}~\bibnamefont {Tatarek}},\ }\bibfield
  {title} {\bibinfo {title} {Anharmonic effects at the onset of the {Ag(110)}
  roughening transition},\ }\href
  {https://doi.org/https://doi.org/10.1016/0039-6028(95)01302-4} {\bibfield
  {journal} {\bibinfo  {journal} {Surf. Sci.}\ }\textbf {\bibinfo {volume}
  {352-354}},\ \bibinfo {pages} {964} (\bibinfo {year} {1996})},\ \bibinfo
  {note} {proceedings of the 15th European Conference on Surface
  Science}\BibitemShut {NoStop}%
\bibitem [{\citenamefont {Novko}\ \emph {et~al.}(2016)\citenamefont {Novko},
  \citenamefont {Alducin}, \citenamefont {Blanco-Rey},\ and\ \citenamefont
  {Juaristi}}]{Novko2016}%
  \BibitemOpen
  \bibfield  {author} {\bibinfo {author} {\bibfnamefont {D.}~\bibnamefont
  {Novko}}, \bibinfo {author} {\bibfnamefont {M.}~\bibnamefont {Alducin}},
  \bibinfo {author} {\bibfnamefont {M.}~\bibnamefont {Blanco-Rey}},\ and\
  \bibinfo {author} {\bibfnamefont {J.~I.}\ \bibnamefont {Juaristi}},\
  }\bibfield  {title} {\bibinfo {title} {Effects of electronic relaxation
  processes on vibrational linewidths of adsorbates on surfaces: The case of
  {CO}/{Cu}(100)},\ }\href {https://doi.org/10.1103/PhysRevB.94.224306}
  {\bibfield  {journal} {\bibinfo  {journal} {Phys. Rev. B}\ }\textbf {\bibinfo
  {volume} {94}},\ \bibinfo {pages} {224306} (\bibinfo {year}
  {2016})}\BibitemShut {NoStop}%
\bibitem [{\citenamefont {Novko}\ \emph {et~al.}(2018)\citenamefont {Novko},
  \citenamefont {Alducin},\ and\ \citenamefont {Juaristi}}]{Novko2018}%
  \BibitemOpen
  \bibfield  {author} {\bibinfo {author} {\bibfnamefont {D.}~\bibnamefont
  {Novko}}, \bibinfo {author} {\bibfnamefont {M.}~\bibnamefont {Alducin}},\
  and\ \bibinfo {author} {\bibfnamefont {J.~I.}\ \bibnamefont {Juaristi}},\
  }\bibfield  {title} {\bibinfo {title} {Electron-mediated phonon-phonon
  coupling drives the vibrational relaxation of {CO} on {Cu}(100)},\ }\href
  {https://doi.org/10.1103/PhysRevLett.120.156804} {\bibfield  {journal}
  {\bibinfo  {journal} {Phys. Rev. Lett.}\ }\textbf {\bibinfo {volume} {120}},\
  \bibinfo {pages} {156804} (\bibinfo {year} {2018})}\BibitemShut {NoStop}%
\bibitem [{\citenamefont {Hayashi}\ \emph {et~al.}(2013)\citenamefont
  {Hayashi}, \citenamefont {Shimada}, \citenamefont {Jiang}, \citenamefont
  {Iwasawa}, \citenamefont {Aiura}, \citenamefont {Oguchi}, \citenamefont
  {Namatame},\ and\ \citenamefont {Taniguchi}}]{Hayashi2013}%
  \BibitemOpen
  \bibfield  {author} {\bibinfo {author} {\bibfnamefont {H.}~\bibnamefont
  {Hayashi}}, \bibinfo {author} {\bibfnamefont {K.}~\bibnamefont {Shimada}},
  \bibinfo {author} {\bibfnamefont {J.}~\bibnamefont {Jiang}}, \bibinfo
  {author} {\bibfnamefont {H.}~\bibnamefont {Iwasawa}}, \bibinfo {author}
  {\bibfnamefont {Y.}~\bibnamefont {Aiura}}, \bibinfo {author} {\bibfnamefont
  {T.}~\bibnamefont {Oguchi}}, \bibinfo {author} {\bibfnamefont
  {H.}~\bibnamefont {Namatame}},\ and\ \bibinfo {author} {\bibfnamefont
  {M.}~\bibnamefont {Taniguchi}},\ }\bibfield  {title} {\bibinfo {title}
  {High-resolution angle-resolved photoemission study of electronic structure
  and electron self-energy in palladium},\ }\href
  {https://doi.org/10.1103/PhysRevB.87.035140} {\bibfield  {journal} {\bibinfo
  {journal} {Phys. Rev. B}\ }\textbf {\bibinfo {volume} {87}},\ \bibinfo
  {pages} {035140} (\bibinfo {year} {2013})}\BibitemShut {NoStop}%
\bibitem [{\citenamefont {Schendel}\ \emph {et~al.}(2017)\citenamefont
  {Schendel}, \citenamefont {Barreteau}, \citenamefont {Brandbyge},
  \citenamefont {Borca}, \citenamefont {Pentegov}, \citenamefont {Schlickum},
  \citenamefont {Ternes}, \citenamefont {Wahl},\ and\ \citenamefont
  {Kern}}]{Schendel2017}%
  \BibitemOpen
  \bibfield  {author} {\bibinfo {author} {\bibfnamefont {V.}~\bibnamefont
  {Schendel}}, \bibinfo {author} {\bibfnamefont {C.}~\bibnamefont {Barreteau}},
  \bibinfo {author} {\bibfnamefont {M.}~\bibnamefont {Brandbyge}}, \bibinfo
  {author} {\bibfnamefont {B.}~\bibnamefont {Borca}}, \bibinfo {author}
  {\bibfnamefont {I.}~\bibnamefont {Pentegov}}, \bibinfo {author}
  {\bibfnamefont {U.}~\bibnamefont {Schlickum}}, \bibinfo {author}
  {\bibfnamefont {M.}~\bibnamefont {Ternes}}, \bibinfo {author} {\bibfnamefont
  {P.}~\bibnamefont {Wahl}},\ and\ \bibinfo {author} {\bibfnamefont
  {K.}~\bibnamefont {Kern}},\ }\bibfield  {title} {\bibinfo {title} {Strong
  paramagnon scattering in single atom {Pd} contacts},\ }\href
  {https://doi.org/10.1103/PhysRevB.96.035155} {\bibfield  {journal} {\bibinfo
  {journal} {Phys. Rev. B}\ }\textbf {\bibinfo {volume} {96}},\ \bibinfo
  {pages} {035155} (\bibinfo {year} {2017})}\BibitemShut {NoStop}%
\bibitem [{\citenamefont {Giustino}(2017)}]{Giustino2017}%
  \BibitemOpen
  \bibfield  {author} {\bibinfo {author} {\bibfnamefont {F.}~\bibnamefont
  {Giustino}},\ }\bibfield  {title} {\bibinfo {title} {Electron-phonon
  interactions from first principles},\ }\href
  {https://doi.org/10.1103/RevModPhys.89.015003} {\bibfield  {journal}
  {\bibinfo  {journal} {Rev. Mod. Phys.}\ }\textbf {\bibinfo {volume} {89}},\
  \bibinfo {pages} {015003} (\bibinfo {year} {2017})}\BibitemShut {NoStop}%
\bibitem [{\citenamefont {Bauer}\ \emph {et~al.}(1998)\citenamefont {Bauer},
  \citenamefont {Schmid}, \citenamefont {Pavone},\ and\ \citenamefont
  {Strauch}}]{Bauer1998}%
  \BibitemOpen
  \bibfield  {author} {\bibinfo {author} {\bibfnamefont {R.}~\bibnamefont
  {Bauer}}, \bibinfo {author} {\bibfnamefont {A.}~\bibnamefont {Schmid}},
  \bibinfo {author} {\bibfnamefont {P.}~\bibnamefont {Pavone}},\ and\ \bibinfo
  {author} {\bibfnamefont {D.}~\bibnamefont {Strauch}},\ }\bibfield  {title}
  {\bibinfo {title} {Electron-phonon coupling in the metallic elements {Al},
  {Au}, {Na}, and {Nb}: A first-principles study},\ }\href
  {https://doi.org/10.1103/PhysRevB.57.11276} {\bibfield  {journal} {\bibinfo
  {journal} {Phys. Rev. B}\ }\textbf {\bibinfo {volume} {57}},\ \bibinfo
  {pages} {11276} (\bibinfo {year} {1998})}\BibitemShut {NoStop}%
\bibitem [{\citenamefont {Anisimov}\ \emph {et~al.}(1974)\citenamefont
  {Anisimov}, \citenamefont {Kapeliovich},\ and\ \citenamefont
  {Perel'Man}}]{Anisimov1974}%
  \BibitemOpen
  \bibfield  {author} {\bibinfo {author} {\bibfnamefont {S.~I.}\ \bibnamefont
  {Anisimov}}, \bibinfo {author} {\bibfnamefont {B.~L.}\ \bibnamefont
  {Kapeliovich}},\ and\ \bibinfo {author} {\bibfnamefont {T.~L.}\ \bibnamefont
  {Perel'Man}},\ }\bibfield  {title} {\bibinfo {title} {Electron emission from
  metal surfaces exposed to ultrashort laser pulses},\ }\href@noop {}
  {\bibfield  {journal} {\bibinfo  {journal} {Sov. Phys. JETP}\ }\textbf
  {\bibinfo {volume} {39}},\ \bibinfo {pages} {375} (\bibinfo {year}
  {1974})}\BibitemShut {NoStop}%
\bibitem [{\citenamefont {Rose}\ \emph {et~al.}(2002)\citenamefont {Rose},
  \citenamefont {Mitsui}, \citenamefont {Dunphy}, \citenamefont {Borg},
  \citenamefont {Ogletree}, \citenamefont {Salmeron},\ and\ \citenamefont
  {Sautet}}]{Rose2002}%
  \BibitemOpen
  \bibfield  {author} {\bibinfo {author} {\bibfnamefont {M.}~\bibnamefont
  {Rose}}, \bibinfo {author} {\bibfnamefont {T.}~\bibnamefont {Mitsui}},
  \bibinfo {author} {\bibfnamefont {J.}~\bibnamefont {Dunphy}}, \bibinfo
  {author} {\bibfnamefont {A.}~\bibnamefont {Borg}}, \bibinfo {author}
  {\bibfnamefont {D.}~\bibnamefont {Ogletree}}, \bibinfo {author}
  {\bibfnamefont {M.}~\bibnamefont {Salmeron}},\ and\ \bibinfo {author}
  {\bibfnamefont {P.}~\bibnamefont {Sautet}},\ }\bibfield  {title} {\bibinfo
  {title} {Ordered structures of {CO} on {Pd}(111) studied by {STM}},\ }\href
  {https://doi.org/https://doi.org/10.1016/S0039-6028(02)01560-1} {\bibfield
  {journal} {\bibinfo  {journal} {Surf. Sci.}\ }\textbf {\bibinfo {volume}
  {512}},\ \bibinfo {pages} {48} (\bibinfo {year} {2002})}\BibitemShut
  {NoStop}%
\bibitem [{\citenamefont {Wellendorff}\ \emph {et~al.}(2012)\citenamefont
  {Wellendorff}, \citenamefont {Lundgaard}, \citenamefont {M\o{}gelh\o{}j},
  \citenamefont {Petzold}, \citenamefont {Landis}, \citenamefont {N\o{}rskov},
  \citenamefont {Bligaard},\ and\ \citenamefont {Jacobsen}}]{Wellendorff2012}%
  \BibitemOpen
  \bibfield  {author} {\bibinfo {author} {\bibfnamefont {J.}~\bibnamefont
  {Wellendorff}}, \bibinfo {author} {\bibfnamefont {K.~T.}\ \bibnamefont
  {Lundgaard}}, \bibinfo {author} {\bibfnamefont {A.}~\bibnamefont
  {M\o{}gelh\o{}j}}, \bibinfo {author} {\bibfnamefont {V.}~\bibnamefont
  {Petzold}}, \bibinfo {author} {\bibfnamefont {D.~D.}\ \bibnamefont {Landis}},
  \bibinfo {author} {\bibfnamefont {J.~K.}\ \bibnamefont {N\o{}rskov}},
  \bibinfo {author} {\bibfnamefont {T.}~\bibnamefont {Bligaard}},\ and\
  \bibinfo {author} {\bibfnamefont {K.~W.}\ \bibnamefont {Jacobsen}},\
  }\bibfield  {title} {\bibinfo {title} {Density functionals for surface
  science: Exchange-correlation model development with bayesian error
  estimation},\ }\href {https://doi.org/10.1103/PhysRevB.85.235149} {\bibfield
  {journal} {\bibinfo  {journal} {Phys. Rev. B}\ }\textbf {\bibinfo {volume}
  {85}},\ \bibinfo {pages} {235149} (\bibinfo {year} {2012})}\BibitemShut
  {NoStop}%
\bibitem [{\citenamefont {Giannozzi}\ \emph {et~al.}(2009)\citenamefont
  {Giannozzi}, \citenamefont {Baroni}, \citenamefont {Bonini}, \citenamefont
  {Calandra}, \citenamefont {Car}, \citenamefont {Cavazzoni}, \citenamefont
  {Ceresoli}, \citenamefont {Chiarotti}, \citenamefont {Cococcioni},
  \citenamefont {Dabo}, \citenamefont {Dal~Corso}, \citenamefont
  {De~Gironcoli}, \citenamefont {Fabris}, \citenamefont {Fratesi},
  \citenamefont {Gebauer}, \citenamefont {Gerstmann}, \citenamefont
  {Gougoussis}, \citenamefont {Kokalj}, \citenamefont {Lazzeri}, \citenamefont
  {Martin-Samos}, \citenamefont {Marzari}, \citenamefont {Mauri}, \citenamefont
  {Mazzarello}, \citenamefont {Paolini}, \citenamefont {Pasquarello},
  \citenamefont {Paulatto}, \citenamefont {Sbraccia}, \citenamefont {Scandolo},
  \citenamefont {Sclauzero}, \citenamefont {Seitsonen}, \citenamefont
  {Smogunov}, \citenamefont {Umari},\ and\ \citenamefont
  {Wentzcovitch}}]{Giannozzi2009}%
  \BibitemOpen
  \bibfield  {author} {\bibinfo {author} {\bibfnamefont {P.}~\bibnamefont
  {Giannozzi}}, \bibinfo {author} {\bibfnamefont {S.}~\bibnamefont {Baroni}},
  \bibinfo {author} {\bibfnamefont {N.}~\bibnamefont {Bonini}}, \bibinfo
  {author} {\bibfnamefont {M.}~\bibnamefont {Calandra}}, \bibinfo {author}
  {\bibfnamefont {R.}~\bibnamefont {Car}}, \bibinfo {author} {\bibfnamefont
  {C.}~\bibnamefont {Cavazzoni}}, \bibinfo {author} {\bibfnamefont
  {D.}~\bibnamefont {Ceresoli}}, \bibinfo {author} {\bibfnamefont {G.~L.}\
  \bibnamefont {Chiarotti}}, \bibinfo {author} {\bibfnamefont {M.}~\bibnamefont
  {Cococcioni}}, \bibinfo {author} {\bibfnamefont {I.}~\bibnamefont {Dabo}},
  \bibinfo {author} {\bibfnamefont {A.}~\bibnamefont {Dal~Corso}}, \bibinfo
  {author} {\bibfnamefont {S.}~\bibnamefont {De~Gironcoli}}, \bibinfo {author}
  {\bibfnamefont {S.}~\bibnamefont {Fabris}}, \bibinfo {author} {\bibfnamefont
  {G.}~\bibnamefont {Fratesi}}, \bibinfo {author} {\bibfnamefont
  {R.}~\bibnamefont {Gebauer}}, \bibinfo {author} {\bibfnamefont
  {U.}~\bibnamefont {Gerstmann}}, \bibinfo {author} {\bibfnamefont
  {C.}~\bibnamefont {Gougoussis}}, \bibinfo {author} {\bibfnamefont
  {A.}~\bibnamefont {Kokalj}}, \bibinfo {author} {\bibfnamefont
  {M.}~\bibnamefont {Lazzeri}}, \bibinfo {author} {\bibfnamefont
  {L.}~\bibnamefont {Martin-Samos}}, \bibinfo {author} {\bibfnamefont
  {N.}~\bibnamefont {Marzari}}, \bibinfo {author} {\bibfnamefont
  {F.}~\bibnamefont {Mauri}}, \bibinfo {author} {\bibfnamefont
  {R.}~\bibnamefont {Mazzarello}}, \bibinfo {author} {\bibfnamefont
  {S.}~\bibnamefont {Paolini}}, \bibinfo {author} {\bibfnamefont
  {A.}~\bibnamefont {Pasquarello}}, \bibinfo {author} {\bibfnamefont
  {L.}~\bibnamefont {Paulatto}}, \bibinfo {author} {\bibfnamefont
  {C.}~\bibnamefont {Sbraccia}}, \bibinfo {author} {\bibfnamefont
  {S.}~\bibnamefont {Scandolo}}, \bibinfo {author} {\bibfnamefont
  {G.}~\bibnamefont {Sclauzero}}, \bibinfo {author} {\bibfnamefont {A.~P.}\
  \bibnamefont {Seitsonen}}, \bibinfo {author} {\bibfnamefont {A.}~\bibnamefont
  {Smogunov}}, \bibinfo {author} {\bibfnamefont {P.}~\bibnamefont {Umari}},\
  and\ \bibinfo {author} {\bibfnamefont {R.~M.}\ \bibnamefont {Wentzcovitch}},\
  }\bibfield  {title} {\bibinfo {title} {{QUANTUM ESPRESSO: A modular and
  open-source software project for quantum simulations of materials}},\ }\href
  {https://doi.org/10.1088/0953-8984/21/39/395502} {\bibfield  {journal}
  {\bibinfo  {journal} {J. Phys. Condens. Matter}\ }\textbf {\bibinfo {volume}
  {21}},\ \bibinfo {pages} {395502} (\bibinfo {year} {2009})}\BibitemShut
  {NoStop}%
\bibitem [{\citenamefont {Giannozzi}\ \emph {et~al.}(2017)\citenamefont
  {Giannozzi}, \citenamefont {Andreussi}, \citenamefont {Brumme}, \citenamefont
  {Bunau}, \citenamefont {{Buongiorno Nardelli}}, \citenamefont {Calandra},
  \citenamefont {Car}, \citenamefont {Cavazzoni}, \citenamefont {Ceresoli},
  \citenamefont {Cococcioni}, \citenamefont {Colonna}, \citenamefont
  {Carnimeo}, \citenamefont {{Dal Corso}}, \citenamefont {{De Gironcoli}},
  \citenamefont {Delugas}, \citenamefont {Distasio}, \citenamefont {Ferretti},
  \citenamefont {Floris}, \citenamefont {Fratesi}, \citenamefont {Fugallo},
  \citenamefont {Gebauer}, \citenamefont {Gerstmann}, \citenamefont {Giustino},
  \citenamefont {Gorni}, \citenamefont {Jia}, \citenamefont {Kawamura},
  \citenamefont {Ko}, \citenamefont {Kokalj}, \citenamefont
  {K{\"{u}}c{\"{u}}kbenli}, \citenamefont {Lazzeri}, \citenamefont {Marsili},
  \citenamefont {Marzari}, \citenamefont {Mauri}, \citenamefont {Nguyen},
  \citenamefont {Nguyen}, \citenamefont {Otero-De-La-Roza}, \citenamefont
  {Paulatto}, \citenamefont {Ponc{\'{e}}}, \citenamefont {Rocca}, \citenamefont
  {Sabatini}, \citenamefont {Santra}, \citenamefont {Schlipf}, \citenamefont
  {Seitsonen}, \citenamefont {Smogunov}, \citenamefont {Timrov}, \citenamefont
  {Thonhauser}, \citenamefont {Umari}, \citenamefont {Vast}, \citenamefont
  {Wu},\ and\ \citenamefont {Baroni}}]{Giannozzi2017}%
  \BibitemOpen
  \bibfield  {author} {\bibinfo {author} {\bibfnamefont {P.}~\bibnamefont
  {Giannozzi}}, \bibinfo {author} {\bibfnamefont {O.}~\bibnamefont
  {Andreussi}}, \bibinfo {author} {\bibfnamefont {T.}~\bibnamefont {Brumme}},
  \bibinfo {author} {\bibfnamefont {O.}~\bibnamefont {Bunau}}, \bibinfo
  {author} {\bibfnamefont {M.}~\bibnamefont {{Buongiorno Nardelli}}}, \bibinfo
  {author} {\bibfnamefont {M.}~\bibnamefont {Calandra}}, \bibinfo {author}
  {\bibfnamefont {R.}~\bibnamefont {Car}}, \bibinfo {author} {\bibfnamefont
  {C.}~\bibnamefont {Cavazzoni}}, \bibinfo {author} {\bibfnamefont
  {D.}~\bibnamefont {Ceresoli}}, \bibinfo {author} {\bibfnamefont
  {M.}~\bibnamefont {Cococcioni}}, \bibinfo {author} {\bibfnamefont
  {N.}~\bibnamefont {Colonna}}, \bibinfo {author} {\bibfnamefont
  {I.}~\bibnamefont {Carnimeo}}, \bibinfo {author} {\bibfnamefont
  {A.}~\bibnamefont {{Dal Corso}}}, \bibinfo {author} {\bibfnamefont
  {S.}~\bibnamefont {{De Gironcoli}}}, \bibinfo {author} {\bibfnamefont
  {P.}~\bibnamefont {Delugas}}, \bibinfo {author} {\bibfnamefont {R.~A.}\
  \bibnamefont {Distasio}}, \bibinfo {author} {\bibfnamefont {A.}~\bibnamefont
  {Ferretti}}, \bibinfo {author} {\bibfnamefont {A.}~\bibnamefont {Floris}},
  \bibinfo {author} {\bibfnamefont {G.}~\bibnamefont {Fratesi}}, \bibinfo
  {author} {\bibfnamefont {G.}~\bibnamefont {Fugallo}}, \bibinfo {author}
  {\bibfnamefont {R.}~\bibnamefont {Gebauer}}, \bibinfo {author} {\bibfnamefont
  {U.}~\bibnamefont {Gerstmann}}, \bibinfo {author} {\bibfnamefont
  {F.}~\bibnamefont {Giustino}}, \bibinfo {author} {\bibfnamefont
  {T.}~\bibnamefont {Gorni}}, \bibinfo {author} {\bibfnamefont
  {J.}~\bibnamefont {Jia}}, \bibinfo {author} {\bibfnamefont {M.}~\bibnamefont
  {Kawamura}}, \bibinfo {author} {\bibfnamefont {H.~Y.}\ \bibnamefont {Ko}},
  \bibinfo {author} {\bibfnamefont {A.}~\bibnamefont {Kokalj}}, \bibinfo
  {author} {\bibfnamefont {E.}~\bibnamefont {K{\"{u}}c{\"{u}}kbenli}}, \bibinfo
  {author} {\bibfnamefont {M.}~\bibnamefont {Lazzeri}}, \bibinfo {author}
  {\bibfnamefont {M.}~\bibnamefont {Marsili}}, \bibinfo {author} {\bibfnamefont
  {N.}~\bibnamefont {Marzari}}, \bibinfo {author} {\bibfnamefont
  {F.}~\bibnamefont {Mauri}}, \bibinfo {author} {\bibfnamefont {N.~L.}\
  \bibnamefont {Nguyen}}, \bibinfo {author} {\bibfnamefont {H.~V.}\
  \bibnamefont {Nguyen}}, \bibinfo {author} {\bibfnamefont {A.}~\bibnamefont
  {Otero-De-La-Roza}}, \bibinfo {author} {\bibfnamefont {L.}~\bibnamefont
  {Paulatto}}, \bibinfo {author} {\bibfnamefont {S.}~\bibnamefont
  {Ponc{\'{e}}}}, \bibinfo {author} {\bibfnamefont {D.}~\bibnamefont {Rocca}},
  \bibinfo {author} {\bibfnamefont {R.}~\bibnamefont {Sabatini}}, \bibinfo
  {author} {\bibfnamefont {B.}~\bibnamefont {Santra}}, \bibinfo {author}
  {\bibfnamefont {M.}~\bibnamefont {Schlipf}}, \bibinfo {author} {\bibfnamefont
  {A.~P.}\ \bibnamefont {Seitsonen}}, \bibinfo {author} {\bibfnamefont
  {A.}~\bibnamefont {Smogunov}}, \bibinfo {author} {\bibfnamefont
  {I.}~\bibnamefont {Timrov}}, \bibinfo {author} {\bibfnamefont
  {T.}~\bibnamefont {Thonhauser}}, \bibinfo {author} {\bibfnamefont
  {P.}~\bibnamefont {Umari}}, \bibinfo {author} {\bibfnamefont
  {N.}~\bibnamefont {Vast}}, \bibinfo {author} {\bibfnamefont {X.}~\bibnamefont
  {Wu}},\ and\ \bibinfo {author} {\bibfnamefont {S.}~\bibnamefont {Baroni}},\
  }\bibfield  {title} {\bibinfo {title} {{Advanced} capabilities for materials
  modelling with {Quantum ESPRESSO}},\ }\href
  {https://doi.org/10.1088/1361-648X/aa8f79} {\bibfield  {journal} {\bibinfo
  {journal} {J. Phys. Condens. Matter}\ }\textbf {\bibinfo {volume} {29}},\
  \bibinfo {pages} {465901} (\bibinfo {year} {2017})}\BibitemShut {NoStop}%
\bibitem [{\citenamefont {Noffsinger}\ \emph {et~al.}(2010)\citenamefont
  {Noffsinger}, \citenamefont {Giustino}, \citenamefont {Malone}, \citenamefont
  {Park}, \citenamefont {Louie},\ and\ \citenamefont
  {Cohen}}]{NOFFSINGER20102140}%
  \BibitemOpen
  \bibfield  {author} {\bibinfo {author} {\bibfnamefont {J.}~\bibnamefont
  {Noffsinger}}, \bibinfo {author} {\bibfnamefont {F.}~\bibnamefont
  {Giustino}}, \bibinfo {author} {\bibfnamefont {B.~D.}\ \bibnamefont
  {Malone}}, \bibinfo {author} {\bibfnamefont {C.-H.}\ \bibnamefont {Park}},
  \bibinfo {author} {\bibfnamefont {S.~G.}\ \bibnamefont {Louie}},\ and\
  \bibinfo {author} {\bibfnamefont {M.~L.}\ \bibnamefont {Cohen}},\ }\bibfield
  {title} {\bibinfo {title} {Epw: A program for calculating the
  electron–phonon coupling using maximally localized {Wannier} functions},\
  }\href {https://doi.org/https://doi.org/10.1016/j.cpc.2010.08.027} {\bibfield
   {journal} {\bibinfo  {journal} {Comput. Phys. Commun.}\ }\textbf {\bibinfo
  {volume} {181}},\ \bibinfo {pages} {2140} (\bibinfo {year}
  {2010})}\BibitemShut {NoStop}%
\bibitem [{\citenamefont {Poncé}\ \emph {et~al.}(2016)\citenamefont {Poncé},
  \citenamefont {Margine}, \citenamefont {Verdi},\ and\ \citenamefont
  {Giustino}}]{PONCE2016116}%
  \BibitemOpen
  \bibfield  {author} {\bibinfo {author} {\bibfnamefont {S.}~\bibnamefont
  {Poncé}}, \bibinfo {author} {\bibfnamefont {E.}~\bibnamefont {Margine}},
  \bibinfo {author} {\bibfnamefont {C.}~\bibnamefont {Verdi}},\ and\ \bibinfo
  {author} {\bibfnamefont {F.}~\bibnamefont {Giustino}},\ }\bibfield  {title}
  {\bibinfo {title} {Epw: Electron–phonon coupling, transport and
  superconducting properties using maximally localized {Wannier} functions},\
  }\href {https://doi.org/https://doi.org/10.1016/j.cpc.2016.07.028} {\bibfield
   {journal} {\bibinfo  {journal} {Comput. Phys. Commun.}\ }\textbf {\bibinfo
  {volume} {209}},\ \bibinfo {pages} {116 } (\bibinfo {year}
  {2016})}\BibitemShut {NoStop}%
\bibitem [{Sup()}]{Suppmat}%
  \BibitemOpen
  \href {http://link.aps.org/supplemental} {}\bibinfo {note} {See Supplemental
  Material at [URL], which additionally includes
  Refs.\,\cite{dalcorso2014,dalcorsolink,Monkhorst1976,Vitale2020,Mostofi2008,Marzari2012,Mostofi2014,Pizzi2020,haynes2016crc,Nakamoto2006,BRADSHAW1978513,Giebel1998,Ashcroft1988,Kittel1986,Johnson1974,Szymanski2007,Hong2016,Li2022,luttinger60,andreatta19}}\BibitemShut
  {NoStop}%
\bibitem [{\citenamefont {Persson}\ and\ \citenamefont
  {Ryberg}(1985)}]{Persson1985}%
  \BibitemOpen
  \bibfield  {author} {\bibinfo {author} {\bibfnamefont {B.~N.~J.}\
  \bibnamefont {Persson}}\ and\ \bibinfo {author} {\bibfnamefont
  {R.}~\bibnamefont {Ryberg}},\ }\bibfield  {title} {\bibinfo {title}
  {Vibrational phase relaxation at surfaces: {CO} on {Ni(111)}},\ }\href
  {https://doi.org/10.1103/PhysRevLett.54.2119} {\bibfield  {journal} {\bibinfo
   {journal} {Phys. Rev. Lett.}\ }\textbf {\bibinfo {volume} {54}},\ \bibinfo
  {pages} {2119} (\bibinfo {year} {1985})}\BibitemShut {NoStop}%
\bibitem [{\citenamefont {C.~Cook}\ \emph {et~al.}(1997)\citenamefont
  {C.~Cook}, \citenamefont {K.~Clowes},\ and\ \citenamefont
  {M.~McCash}}]{Cook1997}%
  \BibitemOpen
  \bibfield  {author} {\bibinfo {author} {\bibfnamefont {J.}~\bibnamefont
  {C.~Cook}}, \bibinfo {author} {\bibfnamefont {S.}~\bibnamefont {K.~Clowes}},\
  and\ \bibinfo {author} {\bibfnamefont {E.}~\bibnamefont {M.~McCash}},\
  }\bibfield  {title} {\bibinfo {title} {Reflection absorption ir studies of
  vibrational energy transfer processes and adsorption energetics},\ }\href
  {https://doi.org/10.1039/A700208D} {\bibfield  {journal} {\bibinfo  {journal}
  {J. Chem. Soc.{,} Faraday Trans.}\ }\textbf {\bibinfo {volume} {93}},\
  \bibinfo {pages} {2315} (\bibinfo {year} {1997})}\BibitemShut {NoStop}%
\bibitem [{\citenamefont {Sky~Zhou}\ \emph {et~al.}(2020)\citenamefont
  {Sky~Zhou}, \citenamefont {Bianco}, \citenamefont {Monacelli}, \citenamefont
  {Errea}, \citenamefont {Mauri},\ and\ \citenamefont
  {Calandra}}]{SkyZhou2020}%
  \BibitemOpen
  \bibfield  {author} {\bibinfo {author} {\bibfnamefont {J.}~\bibnamefont
  {Sky~Zhou}}, \bibinfo {author} {\bibfnamefont {R.}~\bibnamefont {Bianco}},
  \bibinfo {author} {\bibfnamefont {L.}~\bibnamefont {Monacelli}}, \bibinfo
  {author} {\bibfnamefont {I.}~\bibnamefont {Errea}}, \bibinfo {author}
  {\bibfnamefont {F.}~\bibnamefont {Mauri}},\ and\ \bibinfo {author}
  {\bibfnamefont {M.}~\bibnamefont {Calandra}},\ }\bibfield  {title} {\bibinfo
  {title} {Theory of the thickness dependence of the charge density wave
  transition in {1T-TiTe$_2$}},\ }\href
  {https://doi.org/10.1088/2053-1583/abae7a} {\bibfield  {journal} {\bibinfo
  {journal} {2D Mater,}\ }\textbf {\bibinfo {volume} {7}},\ \bibinfo {pages}
  {045032} (\bibinfo {year} {2020})}\BibitemShut {NoStop}%
\bibitem [{\citenamefont {Zhukov}\ \emph {et~al.}(2002)\citenamefont {Zhukov},
  \citenamefont {Aryasetiawan}, \citenamefont {Chulkov},\ and\ \citenamefont
  {Echenique}}]{Zhukov2002}%
  \BibitemOpen
  \bibfield  {author} {\bibinfo {author} {\bibfnamefont {V.~P.}\ \bibnamefont
  {Zhukov}}, \bibinfo {author} {\bibfnamefont {F.}~\bibnamefont
  {Aryasetiawan}}, \bibinfo {author} {\bibfnamefont {E.~V.}\ \bibnamefont
  {Chulkov}},\ and\ \bibinfo {author} {\bibfnamefont {P.~M.}\ \bibnamefont
  {Echenique}},\ }\bibfield  {title} {\bibinfo {title} {Lifetimes of
  quasiparticle excitations in $4d$ transition metals: Scattering theory and
  {LMTO-RPA-GW} approaches},\ }\href
  {https://doi.org/10.1103/PhysRevB.65.115116} {\bibfield  {journal} {\bibinfo
  {journal} {Phys. Rev. B}\ }\textbf {\bibinfo {volume} {65}},\ \bibinfo
  {pages} {115116} (\bibinfo {year} {2002})}\BibitemShut {NoStop}%
\bibitem [{\citenamefont {{Dal Corso}}(2014{\natexlab{a}})}]{dalcorso2014}%
  \BibitemOpen
  \bibfield  {author} {\bibinfo {author} {\bibfnamefont {A.}~\bibnamefont {{Dal
  Corso}}},\ }\bibfield  {title} {\bibinfo {title} {Pseudopotentials periodic
  table: From {H} to {Pu}},\ }\href
  {https://doi.org/https://doi.org/10.1016/j.commatsci.2014.07.043} {\bibfield
  {journal} {\bibinfo  {journal} {Comp. Mater. Sci.}\ }\textbf {\bibinfo
  {volume} {95}},\ \bibinfo {pages} {337} (\bibinfo {year}
  {2014}{\natexlab{a}})}\BibitemShut {NoStop}%
\bibitem [{\citenamefont {{Dal Corso}}(2014{\natexlab{b}})}]{dalcorsolink}%
  \BibitemOpen
  \bibfield  {author} {\bibinfo {author} {\bibfnamefont {A.}~\bibnamefont {{Dal
  Corso}}},\ }\href@noop {} {}\bibinfo {howpublished}
  {\url{https://dalcorso.github.io/pslibrary/}} (\bibinfo {year}
  {2014}{\natexlab{b}})\BibitemShut {NoStop}%
\bibitem [{\citenamefont {Monkhorst}\ and\ \citenamefont
  {Pack}(1976)}]{Monkhorst1976}%
  \BibitemOpen
  \bibfield  {author} {\bibinfo {author} {\bibfnamefont {H.~J.}\ \bibnamefont
  {Monkhorst}}\ and\ \bibinfo {author} {\bibfnamefont {J.~D.}\ \bibnamefont
  {Pack}},\ }\bibfield  {title} {\bibinfo {title} {Special points for
  brillouin-zone integrations},\ }\href
  {https://doi.org/10.1103/PhysRevB.13.5188} {\bibfield  {journal} {\bibinfo
  {journal} {Phys. Rev. B}\ }\textbf {\bibinfo {volume} {13}},\ \bibinfo
  {pages} {5188} (\bibinfo {year} {1976})}\BibitemShut {NoStop}%
\bibitem [{\citenamefont {Vitale}\ \emph {et~al.}(2020)\citenamefont {Vitale},
  \citenamefont {Pizzi}, \citenamefont {Marrazzo}, \citenamefont {Yates},
  \citenamefont {Marzari},\ and\ \citenamefont {Mostofi}}]{Vitale2020}%
  \BibitemOpen
  \bibfield  {author} {\bibinfo {author} {\bibfnamefont {V.}~\bibnamefont
  {Vitale}}, \bibinfo {author} {\bibfnamefont {G.}~\bibnamefont {Pizzi}},
  \bibinfo {author} {\bibfnamefont {A.}~\bibnamefont {Marrazzo}}, \bibinfo
  {author} {\bibfnamefont {J.~R.}\ \bibnamefont {Yates}}, \bibinfo {author}
  {\bibfnamefont {N.}~\bibnamefont {Marzari}},\ and\ \bibinfo {author}
  {\bibfnamefont {A.~A.}\ \bibnamefont {Mostofi}},\ }\bibfield  {title}
  {\bibinfo {title} {Automated high-throughput wannierisation},\ }\href
  {https://doi.org/10.1038/s41524-020-0312-y} {\bibfield  {journal} {\bibinfo
  {journal} {npj Computational Materials}\ }\textbf {\bibinfo {volume} {6}},\
  \bibinfo {pages} {66} (\bibinfo {year} {2020})}\BibitemShut {NoStop}%
\bibitem [{\citenamefont {Mostofi}\ \emph {et~al.}(2008)\citenamefont
  {Mostofi}, \citenamefont {Yates}, \citenamefont {Lee}, \citenamefont {Souza},
  \citenamefont {Vanderbilt},\ and\ \citenamefont {Marzari}}]{Mostofi2008}%
  \BibitemOpen
  \bibfield  {author} {\bibinfo {author} {\bibfnamefont {A.~A.}\ \bibnamefont
  {Mostofi}}, \bibinfo {author} {\bibfnamefont {J.~R.}\ \bibnamefont {Yates}},
  \bibinfo {author} {\bibfnamefont {Y.-S.}\ \bibnamefont {Lee}}, \bibinfo
  {author} {\bibfnamefont {I.}~\bibnamefont {Souza}}, \bibinfo {author}
  {\bibfnamefont {D.}~\bibnamefont {Vanderbilt}},\ and\ \bibinfo {author}
  {\bibfnamefont {N.}~\bibnamefont {Marzari}},\ }\bibfield  {title} {\bibinfo
  {title} {{Wannier90}: A tool for obtaining maximally-localised {Wannier}
  functions},\ }\href
  {https://doi.org/https://doi.org/10.1016/j.cpc.2007.11.016} {\bibfield
  {journal} {\bibinfo  {journal} {Comput. Phys. Commun.}\ }\textbf {\bibinfo
  {volume} {178}},\ \bibinfo {pages} {685 } (\bibinfo {year}
  {2008})}\BibitemShut {NoStop}%
\bibitem [{\citenamefont {Marzari}\ \emph {et~al.}(2012)\citenamefont
  {Marzari}, \citenamefont {Mostofi}, \citenamefont {Yates}, \citenamefont
  {Souza},\ and\ \citenamefont {Vanderbilt}}]{Marzari2012}%
  \BibitemOpen
  \bibfield  {author} {\bibinfo {author} {\bibfnamefont {N.}~\bibnamefont
  {Marzari}}, \bibinfo {author} {\bibfnamefont {A.~A.}\ \bibnamefont
  {Mostofi}}, \bibinfo {author} {\bibfnamefont {J.~R.}\ \bibnamefont {Yates}},
  \bibinfo {author} {\bibfnamefont {I.}~\bibnamefont {Souza}},\ and\ \bibinfo
  {author} {\bibfnamefont {D.}~\bibnamefont {Vanderbilt}},\ }\bibfield  {title}
  {\bibinfo {title} {Maximally localized {Wannier}functions: Theory and
  applications},\ }\href {https://doi.org/10.1103/RevModPhys.84.1419}
  {\bibfield  {journal} {\bibinfo  {journal} {Rev. Mod. Phys.}\ }\textbf
  {\bibinfo {volume} {84}},\ \bibinfo {pages} {1419} (\bibinfo {year}
  {2012})}\BibitemShut {NoStop}%
\bibitem [{\citenamefont {Mostofi}\ \emph {et~al.}(2014)\citenamefont
  {Mostofi}, \citenamefont {Yates}, \citenamefont {Pizzi}, \citenamefont {Lee},
  \citenamefont {Souza}, \citenamefont {Vanderbilt},\ and\ \citenamefont
  {Marzari}}]{Mostofi2014}%
  \BibitemOpen
  \bibfield  {author} {\bibinfo {author} {\bibfnamefont {A.~A.}\ \bibnamefont
  {Mostofi}}, \bibinfo {author} {\bibfnamefont {J.~R.}\ \bibnamefont {Yates}},
  \bibinfo {author} {\bibfnamefont {G.}~\bibnamefont {Pizzi}}, \bibinfo
  {author} {\bibfnamefont {Y.-S.}\ \bibnamefont {Lee}}, \bibinfo {author}
  {\bibfnamefont {I.}~\bibnamefont {Souza}}, \bibinfo {author} {\bibfnamefont
  {D.}~\bibnamefont {Vanderbilt}},\ and\ \bibinfo {author} {\bibfnamefont
  {N.}~\bibnamefont {Marzari}},\ }\bibfield  {title} {\bibinfo {title} {An
  updated version of {Wannier90}: A tool for obtaining maximally-localised
  {Wannier} functions},\ }\href
  {https://doi.org/https://doi.org/10.1016/j.cpc.2014.05.003} {\bibfield
  {journal} {\bibinfo  {journal} {Comput. Phys. Commun.}\ }\textbf {\bibinfo
  {volume} {185}},\ \bibinfo {pages} {2309 } (\bibinfo {year}
  {2014})}\BibitemShut {NoStop}%
\bibitem [{\citenamefont {Pizzi}\ \emph {et~al.}(2020)\citenamefont {Pizzi},
  \citenamefont {Vitale}, \citenamefont {Arita}, \citenamefont {Blügel},
  \citenamefont {Freimuth}, \citenamefont {G{\'{e}}ranton}, \citenamefont
  {Gibertini}, \citenamefont {Gresch}, \citenamefont {Johnson}, \citenamefont
  {Koretsune}, \citenamefont {Iba{\~{n}}ez-Azpiroz}, \citenamefont {Lee},
  \citenamefont {Lihm}, \citenamefont {Marchand}, \citenamefont {Marrazzo},
  \citenamefont {Mokrousov}, \citenamefont {Mustafa}, \citenamefont {Nohara},
  \citenamefont {Nomura}, \citenamefont {Paulatto}, \citenamefont
  {Ponc{\'{e}}}, \citenamefont {Ponweiser}, \citenamefont {Qiao}, \citenamefont
  {Thöle}, \citenamefont {Tsirkin}, \citenamefont {Wierzbowska}, \citenamefont
  {Marzari}, \citenamefont {Vanderbilt}, \citenamefont {Souza}, \citenamefont
  {Mostofi},\ and\ \citenamefont {Yates}}]{Pizzi2020}%
  \BibitemOpen
  \bibfield  {author} {\bibinfo {author} {\bibfnamefont {G.}~\bibnamefont
  {Pizzi}}, \bibinfo {author} {\bibfnamefont {V.}~\bibnamefont {Vitale}},
  \bibinfo {author} {\bibfnamefont {R.}~\bibnamefont {Arita}}, \bibinfo
  {author} {\bibfnamefont {S.}~\bibnamefont {Blügel}}, \bibinfo {author}
  {\bibfnamefont {F.}~\bibnamefont {Freimuth}}, \bibinfo {author}
  {\bibfnamefont {G.}~\bibnamefont {G{\'{e}}ranton}}, \bibinfo {author}
  {\bibfnamefont {M.}~\bibnamefont {Gibertini}}, \bibinfo {author}
  {\bibfnamefont {D.}~\bibnamefont {Gresch}}, \bibinfo {author} {\bibfnamefont
  {C.}~\bibnamefont {Johnson}}, \bibinfo {author} {\bibfnamefont
  {T.}~\bibnamefont {Koretsune}}, \bibinfo {author} {\bibfnamefont
  {J.}~\bibnamefont {Iba{\~{n}}ez-Azpiroz}}, \bibinfo {author} {\bibfnamefont
  {H.}~\bibnamefont {Lee}}, \bibinfo {author} {\bibfnamefont {J.-M.}\
  \bibnamefont {Lihm}}, \bibinfo {author} {\bibfnamefont {D.}~\bibnamefont
  {Marchand}}, \bibinfo {author} {\bibfnamefont {A.}~\bibnamefont {Marrazzo}},
  \bibinfo {author} {\bibfnamefont {Y.}~\bibnamefont {Mokrousov}}, \bibinfo
  {author} {\bibfnamefont {J.~I.}\ \bibnamefont {Mustafa}}, \bibinfo {author}
  {\bibfnamefont {Y.}~\bibnamefont {Nohara}}, \bibinfo {author} {\bibfnamefont
  {Y.}~\bibnamefont {Nomura}}, \bibinfo {author} {\bibfnamefont
  {L.}~\bibnamefont {Paulatto}}, \bibinfo {author} {\bibfnamefont
  {S.}~\bibnamefont {Ponc{\'{e}}}}, \bibinfo {author} {\bibfnamefont
  {T.}~\bibnamefont {Ponweiser}}, \bibinfo {author} {\bibfnamefont
  {J.}~\bibnamefont {Qiao}}, \bibinfo {author} {\bibfnamefont {F.}~\bibnamefont
  {Thöle}}, \bibinfo {author} {\bibfnamefont {S.~S.}\ \bibnamefont {Tsirkin}},
  \bibinfo {author} {\bibfnamefont {M.}~\bibnamefont {Wierzbowska}}, \bibinfo
  {author} {\bibfnamefont {N.}~\bibnamefont {Marzari}}, \bibinfo {author}
  {\bibfnamefont {D.}~\bibnamefont {Vanderbilt}}, \bibinfo {author}
  {\bibfnamefont {I.}~\bibnamefont {Souza}}, \bibinfo {author} {\bibfnamefont
  {A.~A.}\ \bibnamefont {Mostofi}},\ and\ \bibinfo {author} {\bibfnamefont
  {J.~R.}\ \bibnamefont {Yates}},\ }\bibfield  {title} {\bibinfo {title}
  {{Wannier90} as a community code: new features and applications},\ }\href
  {https://doi.org/10.1088/1361-648x/ab51ff} {\bibfield  {journal} {\bibinfo
  {journal} {J. Phys.: Condens. Matter}\ }\textbf {\bibinfo {volume} {32}},\
  \bibinfo {pages} {165902} (\bibinfo {year} {2020})}\BibitemShut {NoStop}%
\bibitem [{\citenamefont {Haynes}\ \emph {et~al.}(2016)\citenamefont {Haynes},
  \citenamefont {Lide},\ and\ \citenamefont {Bruno}}]{haynes2016crc}%
  \BibitemOpen
  \bibfield  {author} {\bibinfo {author} {\bibfnamefont {W.~M.}\ \bibnamefont
  {Haynes}}, \bibinfo {author} {\bibfnamefont {D.~R.}\ \bibnamefont {Lide}},\
  and\ \bibinfo {author} {\bibfnamefont {T.~J.}\ \bibnamefont {Bruno}},\
  }\href@noop {} {\emph {\bibinfo {title} {CRC handbook of chemistry and
  physics}}}\ (\bibinfo  {publisher} {CRC press},\ \bibinfo {address} {New
  York},\ \bibinfo {year} {2016})\BibitemShut {NoStop}%
\bibitem [{\citenamefont {Nakamoto}(2006)}]{Nakamoto2006}%
  \BibitemOpen
  \bibfield  {author} {\bibinfo {author} {\bibfnamefont {K.}~\bibnamefont
  {Nakamoto}},\ }\href
  {https://doi.org/https://doi.org/10.1002/0470027320.s4104} {\emph {\bibinfo
  {title} {Handbook of Vibrational Spectroscopy}}}\ (\bibinfo  {publisher}
  {John Wiley \& Sons, Ltd},\ \bibinfo {year} {2006})\BibitemShut {NoStop}%
\bibitem [{\citenamefont {Bradshaw}\ and\ \citenamefont
  {Hoffmann}(1978)}]{BRADSHAW1978513}%
  \BibitemOpen
  \bibfield  {author} {\bibinfo {author} {\bibfnamefont {A.}~\bibnamefont
  {Bradshaw}}\ and\ \bibinfo {author} {\bibfnamefont {F.}~\bibnamefont
  {Hoffmann}},\ }\bibfield  {title} {\bibinfo {title} {The chemisorption of
  carbon monoxide on palladium single crystal surfaces: Ir spectroscopic
  evidence for localised site adsorption},\ }\href
  {https://doi.org/https://doi.org/10.1016/0039-6028(78)90367-9} {\bibfield
  {journal} {\bibinfo  {journal} {Surf. Sci.}\ }\textbf {\bibinfo {volume}
  {72}},\ \bibinfo {pages} {513 } (\bibinfo {year} {1978})}\BibitemShut
  {NoStop}%
\bibitem [{\citenamefont {Gießel}\ \emph {et~al.}(1998)\citenamefont
  {Gießel}, \citenamefont {Schaff}, \citenamefont {Hirschmugl}, \citenamefont
  {Fernandez}, \citenamefont {Schindler}, \citenamefont {Theobald},
  \citenamefont {Bao}, \citenamefont {Lindsay}, \citenamefont {Berndt},
  \citenamefont {Bradshaw}, \citenamefont {Baddeley}, \citenamefont {Lee},
  \citenamefont {Lambert},\ and\ \citenamefont {Woodruff}}]{Giebel1998}%
  \BibitemOpen
  \bibfield  {author} {\bibinfo {author} {\bibfnamefont {T.}~\bibnamefont
  {Gießel}}, \bibinfo {author} {\bibfnamefont {O.}~\bibnamefont {Schaff}},
  \bibinfo {author} {\bibfnamefont {C.}~\bibnamefont {Hirschmugl}}, \bibinfo
  {author} {\bibfnamefont {V.}~\bibnamefont {Fernandez}}, \bibinfo {author}
  {\bibfnamefont {K.-M.}\ \bibnamefont {Schindler}}, \bibinfo {author}
  {\bibfnamefont {A.}~\bibnamefont {Theobald}}, \bibinfo {author}
  {\bibfnamefont {S.}~\bibnamefont {Bao}}, \bibinfo {author} {\bibfnamefont
  {R.}~\bibnamefont {Lindsay}}, \bibinfo {author} {\bibfnamefont
  {W.}~\bibnamefont {Berndt}}, \bibinfo {author} {\bibfnamefont
  {A.}~\bibnamefont {Bradshaw}}, \bibinfo {author} {\bibfnamefont
  {C.}~\bibnamefont {Baddeley}}, \bibinfo {author} {\bibfnamefont
  {A.}~\bibnamefont {Lee}}, \bibinfo {author} {\bibfnamefont {R.}~\bibnamefont
  {Lambert}},\ and\ \bibinfo {author} {\bibfnamefont {D.}~\bibnamefont
  {Woodruff}},\ }\bibfield  {title} {\bibinfo {title} {A photoelectron
  diffraction study of ordered structures in the chemisorption system
  {Pd}(111)-{CO}},\ }\href
  {https://doi.org/https://doi.org/10.1016/S0039-6028(98)00098-3} {\bibfield
  {journal} {\bibinfo  {journal} {Surf. Sci.}\ }\textbf {\bibinfo {volume}
  {406}},\ \bibinfo {pages} {90 } (\bibinfo {year} {1998})}\BibitemShut
  {NoStop}%
\bibitem [{\citenamefont {Ashcroft}\ and\ \citenamefont
  {Mermin}(1988)}]{Ashcroft1988}%
  \BibitemOpen
  \bibfield  {author} {\bibinfo {author} {\bibfnamefont {N.~W.}\ \bibnamefont
  {Ashcroft}}\ and\ \bibinfo {author} {\bibfnamefont {N.~D.}\ \bibnamefont
  {Mermin}},\ }\href@noop {} {\emph {\bibinfo {title} {Solid state physics}}},\
  edited by\ \bibinfo {editor} {\bibfnamefont {P.}~\bibnamefont
  {Saunders~College}}\ (\bibinfo {year} {1988})\BibitemShut {NoStop}%
\bibitem [{\citenamefont {Kittel}(1986)}]{Kittel1986}%
  \BibitemOpen
  \bibfield  {author} {\bibinfo {author} {\bibfnamefont {C.}~\bibnamefont
  {Kittel}},\ }\href@noop {} {\emph {\bibinfo {title} {Introduction to solid
  state physics}}},\ \bibinfo {edition} {6th}\ ed.,\ edited by\ \bibinfo
  {editor} {\bibfnamefont {N.~Y.}\ \bibnamefont {Wiley \&~Sons}}\ (\bibinfo
  {year} {1986})\BibitemShut {NoStop}%
\bibitem [{\citenamefont {Johnson}\ and\ \citenamefont
  {Christy}()}]{Johnson1974}%
  \BibitemOpen
  \bibfield  {author} {\bibinfo {author} {\bibfnamefont {P.}~\bibnamefont
  {Johnson}}\ and\ \bibinfo {author} {\bibfnamefont {R.}~\bibnamefont
  {Christy}},\ }\bibfield  {title} {\bibinfo {title} {Optical constants of
  transition metals: {Ti, V, Cr, Mn, Fe, Co, Ni, and Pd}},\ }\href
  {https://doi.org/10.1103/PhysRevB.9.5056} {\bibfield  {journal} {\bibinfo
  {journal} {Phys. Rev. B}\ }\textbf {\bibinfo {volume} {9}},\ \bibinfo {pages}
  {5056}}\BibitemShut {NoStop}%
\bibitem [{\citenamefont {Szymanski}\ \emph {et~al.}(2007)\citenamefont
  {Szymanski}, \citenamefont {Harris},\ and\ \citenamefont
  {Camillone}}]{Szymanski2007}%
  \BibitemOpen
  \bibfield  {author} {\bibinfo {author} {\bibfnamefont {P.}~\bibnamefont
  {Szymanski}}, \bibinfo {author} {\bibfnamefont {A.}~\bibnamefont {Harris}},\
  and\ \bibinfo {author} {\bibfnamefont {N.}~\bibnamefont {Camillone}},\
  }\bibfield  {title} {\bibinfo {title} {Temperature-dependent
  electron-mediated coupling in subpicosecond photoinduced desorption},\ }\href
  {https://doi.org/https://doi.org/10.1016/j.susc.2007.06.004} {\bibfield
  {journal} {\bibinfo  {journal} {Surf. Sci.}\ }\textbf {\bibinfo {volume}
  {601}},\ \bibinfo {pages} {3335} (\bibinfo {year} {2007})}\BibitemShut
  {NoStop}%
\bibitem [{\citenamefont {Hong}\ \emph {et~al.}(2016)\citenamefont {Hong},
  \citenamefont {Xu}, \citenamefont {Camillone}, \citenamefont {White},\ and\
  \citenamefont {Camillone}}]{Hong2016}%
  \BibitemOpen
  \bibfield  {author} {\bibinfo {author} {\bibfnamefont {S.-Y.}\ \bibnamefont
  {Hong}}, \bibinfo {author} {\bibfnamefont {P.}~\bibnamefont {Xu}}, \bibinfo
  {author} {\bibfnamefont {N.~R.}\ \bibnamefont {Camillone}}, \bibinfo {author}
  {\bibfnamefont {M.~G.}\ \bibnamefont {White}},\ and\ \bibinfo {author}
  {\bibfnamefont {N.}~\bibnamefont {Camillone}},\ }\bibfield  {title} {\bibinfo
  {title} {Adlayer structure dependent ultrafast desorption dynamics in carbon
  monoxide adsorbed on {Pd} (111)},\ }\href {https://doi.org/10.1063/1.4954408}
  {\bibfield  {journal} {\bibinfo  {journal} {J. Chem. Phys.}\ }\textbf
  {\bibinfo {volume} {145}},\ \bibinfo {pages} {014704} (\bibinfo {year}
  {2016})}\BibitemShut {NoStop}%
\bibitem [{\citenamefont {Li}\ and\ \citenamefont {Ji}(2022)}]{Li2022}%
  \BibitemOpen
  \bibfield  {author} {\bibinfo {author} {\bibfnamefont {Y.}~\bibnamefont
  {Li}}\ and\ \bibinfo {author} {\bibfnamefont {P.}~\bibnamefont {Ji}},\
  }\bibfield  {title} {\bibinfo {title} {Ab initio calculation of electron
  temperature dependent electron heat capacity and electron-phonon coupling
  factor of noble metals},\ }\href
  {https://doi.org/https://doi.org/10.1016/j.commatsci.2021.110959} {\bibfield
  {journal} {\bibinfo  {journal} {Comp. Mater Sci.}\ }\textbf {\bibinfo
  {volume} {202}},\ \bibinfo {pages} {110959} (\bibinfo {year}
  {2022})}\BibitemShut {NoStop}%
\bibitem [{\citenamefont {Luttinger}(1960)}]{luttinger60}%
  \BibitemOpen
  \bibfield  {author} {\bibinfo {author} {\bibfnamefont {J.~M.}\ \bibnamefont
  {Luttinger}},\ }\bibfield  {title} {\bibinfo {title} {Fermi surface and some
  simple equilibrium properties of a system of interacting fermions},\ }\href
  {https://doi.org/10.1103/PhysRev.119.1153} {\bibfield  {journal} {\bibinfo
  {journal} {Phys. Rev.}\ }\textbf {\bibinfo {volume} {119}},\ \bibinfo {pages}
  {1153} (\bibinfo {year} {1960})}\BibitemShut {NoStop}%
\bibitem [{\citenamefont {Andreatta}\ \emph {et~al.}(2019)\citenamefont
  {Andreatta}, \citenamefont {Rostami}, \citenamefont {\ifmmode~\check{C}\else
  \v{C}\fi{}abo}, \citenamefont {Bianchi}, \citenamefont {Sanders},
  \citenamefont {Biswas}, \citenamefont {Cacho}, \citenamefont {Jones},
  \citenamefont {Chapman}, \citenamefont {Springate}, \citenamefont {King},
  \citenamefont {Miwa}, \citenamefont {Balatsky}, \citenamefont {Ulstrup},\
  and\ \citenamefont {Hofmann}}]{andreatta19}%
  \BibitemOpen
  \bibfield  {author} {\bibinfo {author} {\bibfnamefont {F.}~\bibnamefont
  {Andreatta}}, \bibinfo {author} {\bibfnamefont {H.}~\bibnamefont {Rostami}},
  \bibinfo {author} {\bibfnamefont {A.~G.}\ \bibnamefont
  {\ifmmode~\check{C}\else \v{C}\fi{}abo}}, \bibinfo {author} {\bibfnamefont
  {M.}~\bibnamefont {Bianchi}}, \bibinfo {author} {\bibfnamefont {C.~E.}\
  \bibnamefont {Sanders}}, \bibinfo {author} {\bibfnamefont {D.}~\bibnamefont
  {Biswas}}, \bibinfo {author} {\bibfnamefont {C.}~\bibnamefont {Cacho}},
  \bibinfo {author} {\bibfnamefont {A.~J.~H.}\ \bibnamefont {Jones}}, \bibinfo
  {author} {\bibfnamefont {R.~T.}\ \bibnamefont {Chapman}}, \bibinfo {author}
  {\bibfnamefont {E.}~\bibnamefont {Springate}}, \bibinfo {author}
  {\bibfnamefont {P.~D.~C.}\ \bibnamefont {King}}, \bibinfo {author}
  {\bibfnamefont {J.~A.}\ \bibnamefont {Miwa}}, \bibinfo {author}
  {\bibfnamefont {A.}~\bibnamefont {Balatsky}}, \bibinfo {author}
  {\bibfnamefont {S.}~\bibnamefont {Ulstrup}},\ and\ \bibinfo {author}
  {\bibfnamefont {P.}~\bibnamefont {Hofmann}},\ }\bibfield  {title} {\bibinfo
  {title} {Transient hot electron dynamics in single-layer
  {${\mathrm{TaS}}_{2}$}},\ }\href {https://doi.org/10.1103/PhysRevB.99.165421}
  {\bibfield  {journal} {\bibinfo  {journal} {Phys. Rev. B}\ }\textbf {\bibinfo
  {volume} {99}},\ \bibinfo {pages} {165421} (\bibinfo {year}
  {2019})}\BibitemShut {NoStop}%
\end{thebibliography}%

\end{document}